\documentclass{aa}

\usepackage{graphicx}
\usepackage{txfonts}

\usepackage[normalem]{ulem}

\usepackage{hyperref}
\hypersetup{
    colorlinks=true,
    linkcolor=blue,
    citecolor = blue,
    urlcolor=black
    }

\begin{document}

\title{Loose threads: parsec-scale filamentation in the high Galactic latitude molecular clouds MBM 3 and MBM 16}

\author{Marco Monaci
          \inst{1,5}\orcid{0000-0003-1319-7714}
          \and
          Loris Magnani
          \inst{2}\orcid{0000-0002-6365-7320}
          \and
          Steven N.Shore
          \inst{1,3,4}\orcid{0000-0003-1677-8004}
          }

\institute{Dipartimento di Fisica, Università di Pisa, Largo Bruno Pontecorvo 3, Pisa, Italy \\
            \email{steven.neil.shore@unipi.it} \\
            \email{mmonaci@swin.edu.au}
        \and
            Department of Physics and Astronomy, University of Georgia, Athens, GA 30602-2451 USA\\
            \email{loris@uga.edu}            
        \and
             INAF-OATS, Via G.B. Tiepolo 11, 34143 Trieste, Italy
        \and
             INFN - Sezione di Pisa, largo B. Pontecorvo 3, Pisa 56127 Italy
        \and
            Centre for Astrophysics and Supercomputing, Swinburne University, John Street, Hawthorn, VIC 3122 Australia}

\date{Received: -; accepted: - }

  \abstract
  % context heading (optional)
   {The existence of high galactic latitude molecular clouds has been known for several decades, and studies of their dust and gas distributions reveal complicated morphological structures. Their dynamics involve turbulence even in the absence of internal energy sources such as stars.}
  % aims heading (mandatory)
   {We study in detail two such clouds, MBM 3 and MBM 16, trying to recover the geometric structure and topology of the gas distribution. In particular, we address the evidence of superthermal asymmetric atomic and molecular line profiles as a result of filament superposition combined with turbulent motions.}
  % methods heading (mandatory)
   {We use a variety of spectroscopic and imaging archival observations of the gas and dust components. The spectroscopic data set comprises \ion{H}{i} 21 cm, \element[][12]CO, \element[][13]CO, and CH line profiles. We also use archival infrared images to study the dust distribution and temperature. To understand the topology of MBM 3 and MBM 16 we compare molecular and atomic spectra, along with profile decomposition of the \ion{H}{i} 21 cm line. Standard tools such as Structure Functions of velocity centroids are used to characterise the turbulence in MBM 3, and channel maps and position-velocity diagrams are employed for elucidating the filament topology of both clouds.}
  % results heading (mandatory)
{The unusually large linewidths previously reported for MBM 3 are due to superposition of individual filaments whose superthermal linewidths are about 1 km s$^{-1}$.  In MBM 16, the cloud appears to decompose into two adjacent structures with similar properties.  The filaments have a high aspect ratio, with lengths of about 1 pc and widths of about 0.1 pc.  In general, the molecular gas is embedded within more extended neutral hydrogen structures.   Velocity gradients found within these structures are not necessarily dynamical, convergent flows. Projection effects and topology of the driving flows produce signatures that mimic velocity shears even if they are simply distortions of ordered gas.  }
  % conclusions heading (optional)
  {}


\titlerunning{Filamentation in high latitude molecular clouds}
\maketitle

\section{Introduction}
The Galactic interstellar medium (ISM) displays structural complexity at almost all length scales, from $\sim$100 pc filaments \citep{2016A&A...590A.131A} to small parsec-scale molecular clouds (\citealt{2017ASSL..442.....M}; \citealt{2018A&A...610A..77H}). Both atomic and molecular species show broadened, asymmetric line profiles along individual lines of sight, indicating a turbulent environment (\citealt{2004ARA&A..42..211E}; \citealt{2004ARA&A..42..275S}; \citealt{1985ApJ...295..466K}; \citealt{1985ApJ...295..479D}) and complex two and three-dimensional topology (\citealt{2023A&A...676A.138M}; \citealt{2023ApJ...948..109K}). The discovery of the infrared cirrus, especially notable at high Galactic latitudes \citep[see][]{1984ApJ...278L..19L}, clearly indicated that filaments and sheets are prevalent in the diffuse ISM. Filaments are found in different environments, including non-star forming clouds \citep[see][]{2010A&A...518L.103M, 2023A&A...676A.138M}, low-mass star-forming clouds \citep[see][]{2010A&A...518L.103M, 2011A&A...529L...6A} and even higher-mass clouds \citep[see][]{2011A&A...533A..94H, 2023A&A...674A.225L}. 

Filaments have been studied theoretically by, for example,  \cite{2012A&A...542A..77F}, who considered isothermal self-gravitating infinitely long cylinders, and \cite{2016MNRAS.457..375F}, who performed a 3D high-resolution magnetohydrodynamical simulation of molecular clouds that form filaments and stars.  The latter found that $\sim$ 0.1 pc-width filaments are remarkably universal, consistent with  observations \citep[see][]{2011A&A...529L...6A, 2012A&A...544A..50M, 2015MNRAS.453.2036B}. The direction of these filaments often seems to be perpendicular to the magnetic field \citep[see][]{2011Natur.478..214G, 2013A&A...556A.153H}. More recently, \cite{2024A&A...686A.155C} analyzed cloud properties of different types in MHD simulations (from small-scale structures such as patches of ISM to full galactic disks) using different codes, finding that the mass spectra seem to be universal, as are filamentary structures. However, in their conclusions, Colman et al. pointed out that the low-mass end of the distribution is highly sensitive to the refinement criterion applied in the various codes; thus, the study of low-mass, diffuse, molecular clouds such as the diffuse and translucent molecular clouds identified at high Galactic latitudes by \cite{1985ApJ...295..402M} is important for constraining models and for understanding filamentation as a product of turbulence in these low mass structures. 

In our previous paper, \cite{2023A&A...676A.138M}, we studied a prototypical High Latitude Molecular Cloud (HLMC), MBM 40, for which we had a significant 
amount of spectra and imaging data for molecular, atomic, and dust components.
We proposed a framework for the recovery of the velocity structure and topology of the gas.  Our intent here is to perform a compare and contrast study of two other HLMCs, MBM 3 and MBM 16, for which substantial and similar data are available.
As for MBM 40, we are trying to recover the 3D geometric structure and topology of the gas distribution of these objects.  In particular, we analyze how the superthermal asymmetric atomic and molecular line profiles can be interpreted as resulting from filament superposition combined with turbulent motions.

\section{The molecular clouds in this study}

Here we describe the principal gas and dust properties of the two clouds in this study as determined from previous \element[][12]CO observations and dust imaging data.

\subsection{MBM 3}

MBM 3 is a HLMC centered at ($\alpha$:$\delta$:J2000.0) $\sim$ (1$^\mathrm{h}$16$^\mathrm{m}$:$16\degr 33\arcmin)$; ($\ell$:$b$) $\sim$ (131.4$^\circ$:$-45.9^\circ$), as obtained from \element[][12]CO observations. The cloud was discovered by \cite{1985ApJ...295..402M} and a complete CO(1-0) map and discussion of the cloud kinematics, energetics, and turbulence characteristics were presented by \cite{2006A&A...457..197S}. There is no evidence for internal star formation \citep[see][]{2005PhDT.......331C}, implying that its internal dynamics are not self-generated. 

 \cite{2006A&A...457..197S}, assuming a distance of about 130 pc, derived a molecular mass of 20-40 M$_{\odot}$ and atomic mass of 10-15 M$_{\odot}$. However, based on reddening derived from subsequent photometric surveys, \cite{2014ApJ...786...29S} revised the distance to 277$_{-26}^{+22}$ pc, which is supported by more recent determinations by \cite{zucker2019}: 314 $\pm$ 15 pc, and \cite{2021ApJS..256...46S}: 309 $\pm$ 4 pc. In this paper, we will adopt  300 pc as the value for the cloud's distance. This new distance yields updated molecular and atomic masses using \element[][12]CO J=(1-0) for the H$_2$ component and the neutral hydrogen 21 cm transition for the atomic component. Adopting the standard relations $N(\ion{H}{i}) =1.813 \cdot 10^{18} \int T_{B,\ion{H}{i}} \ d\mathrm{v} \ cm^{-2}$ \citep[see, e.g.,][]{2011piim.book.....D} and N(H$_2$) $= 2\cdot 10^{20} \int T_{mb,CO} \ d\mathrm{v} \ cm^{-2}$ \citep[see][]{2013ARA&A..51..207B} yields $M_{\mathrm{H_2}} \simeq 80 \ M_{\odot}$ and $M_{\ion{H}{i}} \simeq 20 \ M_{\odot}$. Correcting for the presence of helium, which contributes another 25\%, the total mass of the cloud is M$_{\mathrm{TOT}} \simeq 125  \ M_{\odot}$.  For further discussion, see Appendix \ref{app:mass}. In contrast, MBM 40 has $M_{\mathrm{H_2}} \simeq$ 20-40 M$_{\odot}$ (\citealt{2023A&A...676A.138M}).

\begin{figure*}
  \centering
  \includegraphics[width=18cm]{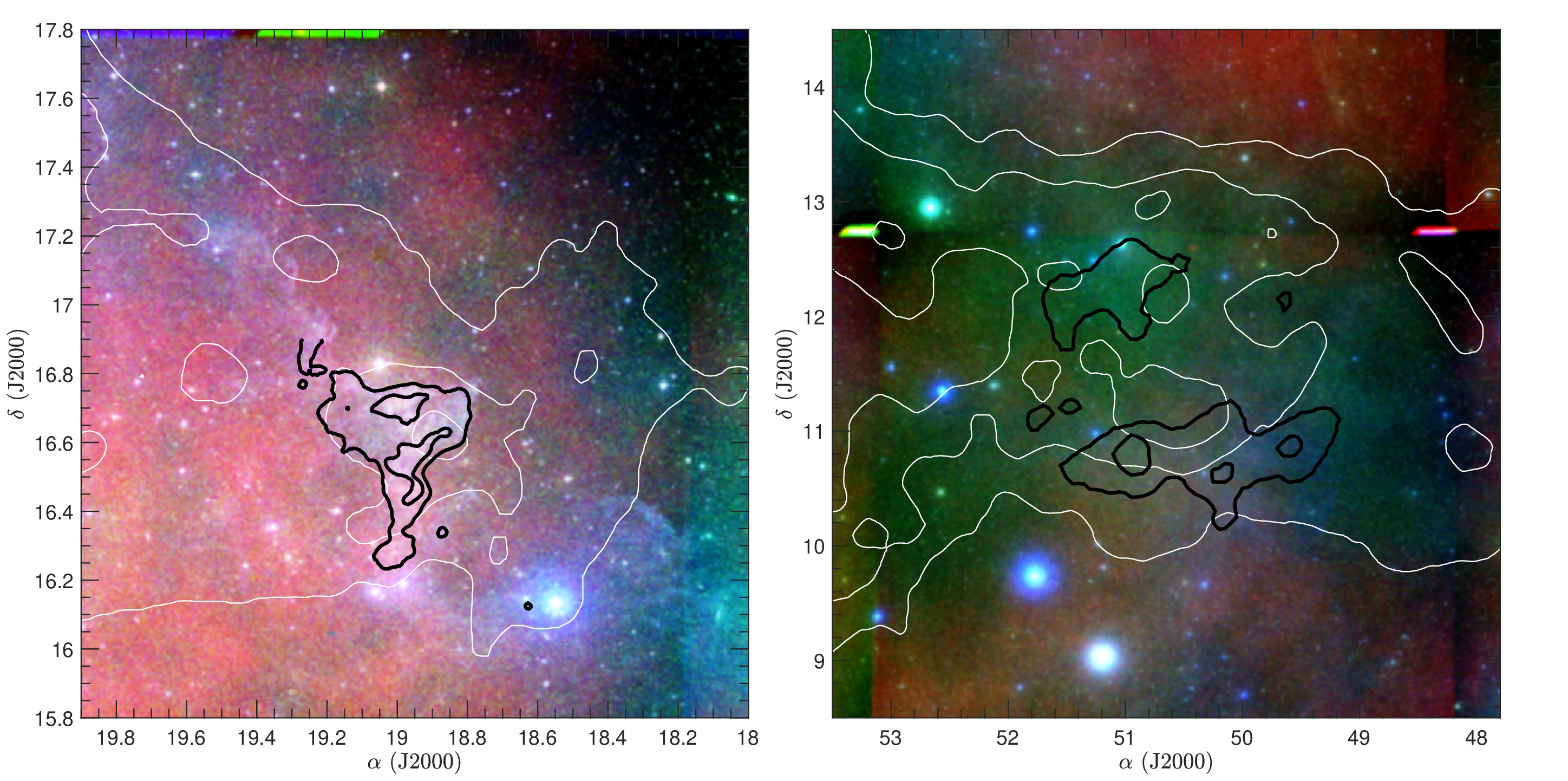}
    \caption{Visible composite image of MBM 03 (left panel) and MBM 16 (right panel). Visible image from DSS2 (red channel = IR filter, green channel = R filter, blue channel = B filter). Black thick contours depict the \element[][12]CO at levels of 4 and 9 K km s$^{-1}$ for MBM 03, and 2 and 4 K km s$^{-1}$ for MBM 16. White thin contours portray the \ion{H}{i} column density at levels of 100, 120, 140 K km s$^{-1}$ for MBM 03, and 230, 270, 310 K km s$^{-1}$ for MBM 16. \element[][12]CO and \ion{H}{i} integrated intensities are calculated in the velocity range of [$-$10.7, $-$3.6] km s$^{-1}$ for MBM 03, and [4.7, 9.3] km s$^{-1}$ for MBM 16.}
        \label{fig:optic_hi_co}
\end{figure*}

Although MBM 3 was previously classified as a translucent cloud,
\cite{1998ApJ...500..525S} derived a maximum E(B-V) of 0.23 mag. Using the total-to-selective extinction ratio $A_V/E(B-V) =$  3.1, the maximum visual extinction is thus $\simeq$ 0.7 mag, classifying MBM 3 as a diffuse cloud following \cite{1988ApJ...334..771V}. The left panel of Figure \ref{fig:optic_hi_co} shows a composite view of MBM 3 in dust and gas (atomic and molecular). The dust reflection very closely matches the molecular gas traced by \element[][12]CO (black thick contours). It is important to note that MBM 3 is a relatively dense molecular knot in a much larger \ion{H}{i} filamentary structure, portions of which are molecular and also associated with HLMCs MBM 4 and DIR 121-45 \citep{1998ApJ...507..507R}. Unlike MBM 40 \citep{2022A&A...668L...9M,2023A&A...676A.138M}, in MBM 3 the atomic hydrogen (depicted in white thin contours in Figure \ref{fig:optic_hi_co}) does not show any particular spatial correlation with dust and molecular gas, even when the selected velocity is the same as the bulk velocity determined from \element[][12]CO observations. We will return to this in Section \ref{subsec:hi}.
The upper (northern) part of the cloud ($16.5 \degr < \delta < 16.8 \degr$) is stronger in emission in both \element[][12]CO and \ion{H}{i}, whereas the lower (southern) part ($16.2 \degr < \delta < 16.5 \degr$) shows weaker emission in both tracers. We will refer to the northern region as the \textit{anvil} and the southern one as the \textit{stem}, due to their respective shapes. 

\begin{figure}
  \centering
  \includegraphics[width=\hsize]{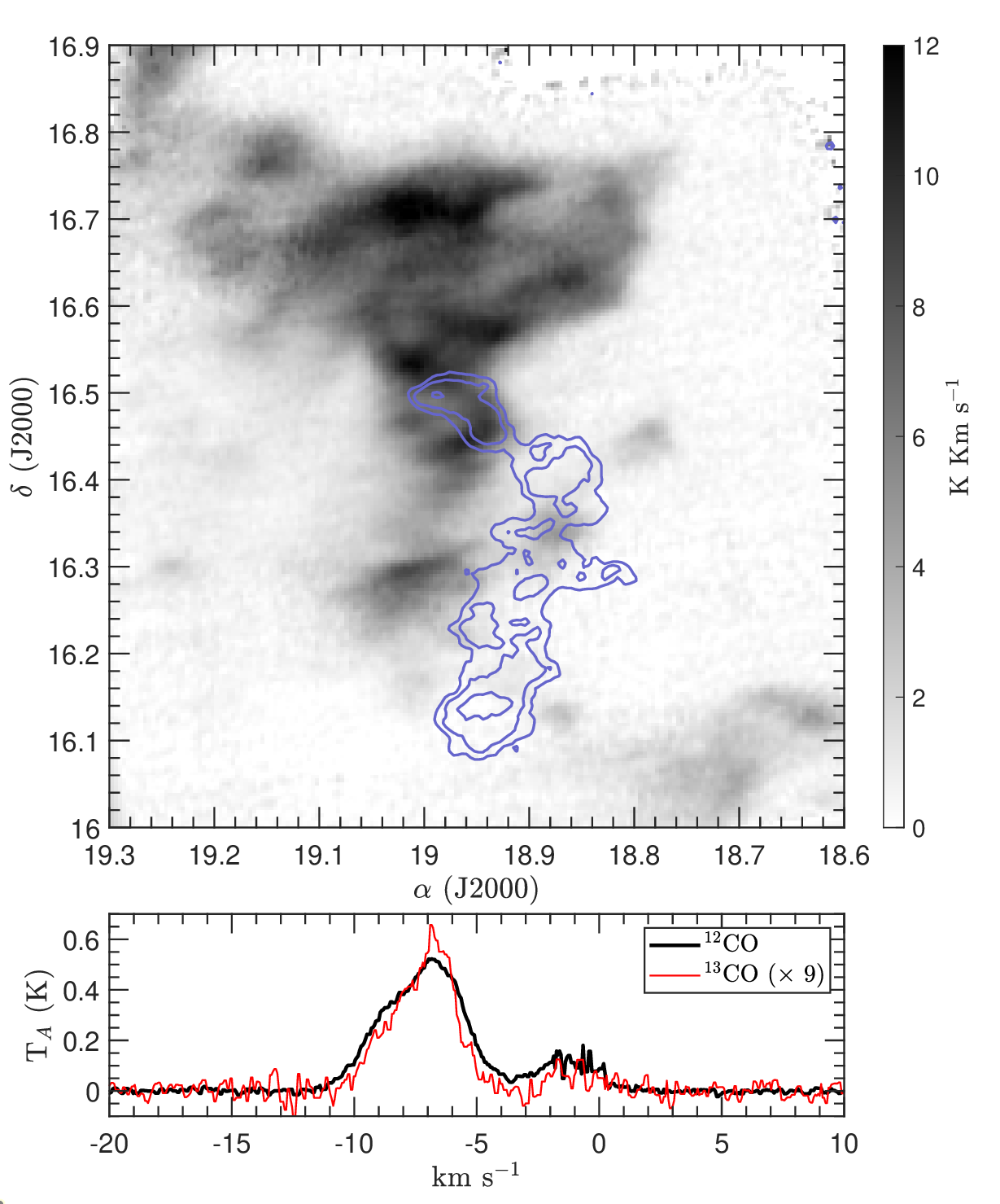}
    \caption{Integrated \element[][12]CO antenna temperature map of MBM 3 of the primary cloud between $-$10.9 km s$^{-1}$ and $-$4.9 km s$^{-1}$ (gray scale), and integrated \element[][12]CO antenna temperature of the outlier cloud between $-$2.5 km s$^{-1}$ and 1.2 km s$^{-1}$ (pale blue contours). The bottom panel shows \element[][12]CO (black thick line) and \element[][13]CO (red thin line) averaged spectrum over the whole cloud. The highest peak traces the gas linked with the primary cloud, while the other peak traces the outlier cloud, which shows virtually no \element[][13]CO emission. The \element[][13]CO spectrum is multiplied by a factor of 9 for clarity.}
        \label{fig:secondary_cloud}
\end{figure}

 In \element[][12]CO(J=1-0) MBM 3 shows two distinct velocity components, one at $\sim$ $-$7 km s$^{-1}$ (LSR) with an unusually large linewidth of about 5 km s$^{-1},${\footnote {High latitude molecular clouds have \element[][12]CO(J=1-0) linewidths typically in the 1-2 km s$^{-1}$ range \citep[see][]{1985ApJ...295..402M}.}} and another at $\sim$ $-$1 km s$^{-1}$. A map of the integrated antenna temperature shows two separate clouds, see the top panel of Figure \ref{fig:secondary_cloud}.  Hereafter, we refer to the cloud at $\sim$ $-$1 km s$^{-1}$ as "the outlier cloud". While it is not clear if the two clouds are physically linked, the dearth of HLMCs at $\vert b \vert \ge$ 30$^\circ$, makes it unlikely that the two structures are completely unrelated \citep[see][]{2017ASSL..442.....M}.  They may have formed within a much larger structure but still be physically and dynamically distinct. 

\subsection{MBM 16}

\begin{figure}
  \centering
  \includegraphics[width=\hsize]{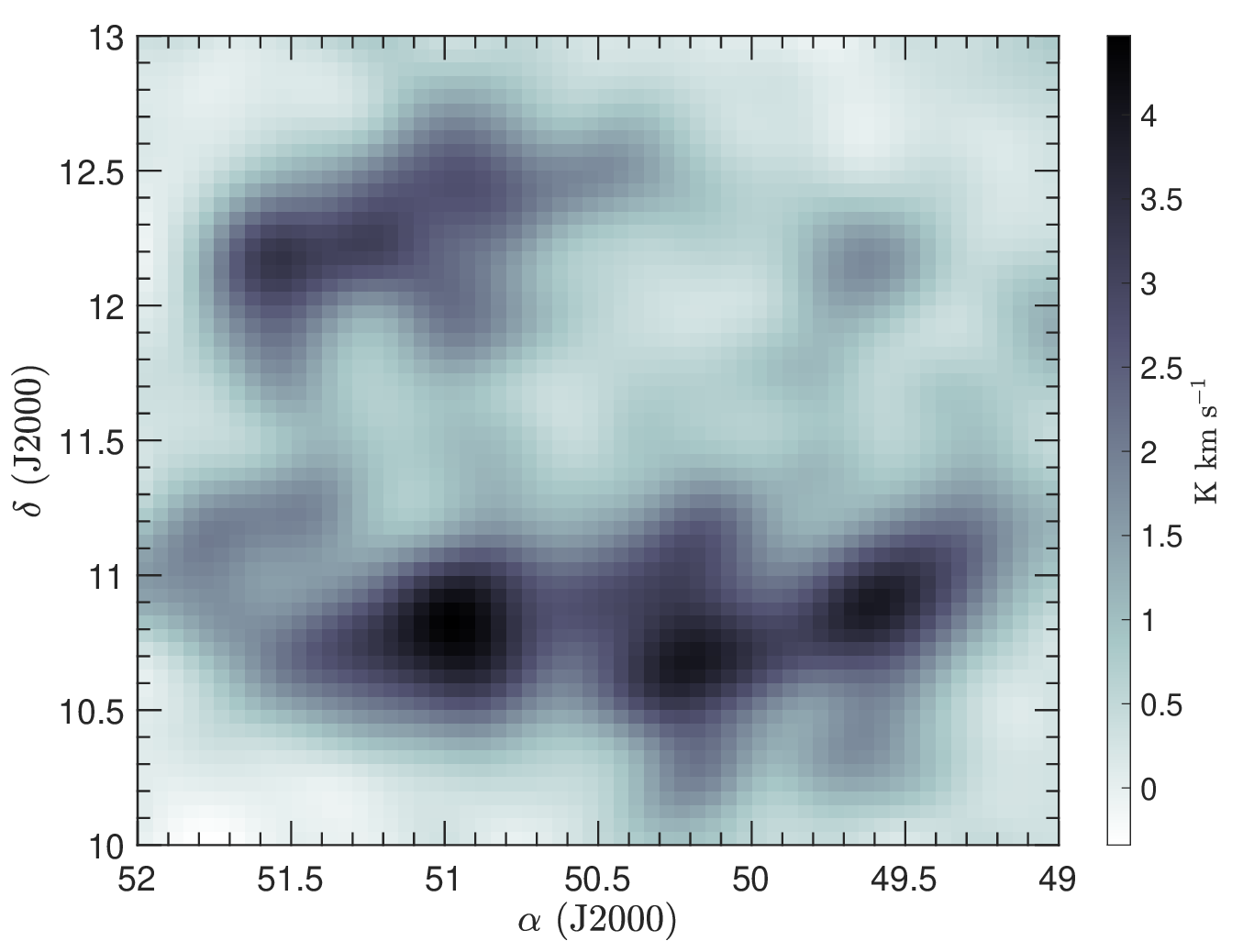}
    \caption{Integrated \element[][12]CO antenna temperature of MBM 16. One can readily see an upper (northern) and lower (southern) filament. The northern branch globally shows a lower antenna temperature and has a detached knot at $\alpha \sim 49.6\degr$. The southern branch presents three knots aligned approximately at the same declination with higher antenna temperatures than the northern branch.}
        \label{fig:mbm16int}
\end{figure}

MBM 16 is an HLMC near the Taurus Dark Cloud complex. The distance, inferred by \cite{zucker2019}, is about 170 pc. The cloud was first detected by \cite{1985ApJ...295..402M} and then fully mapped  in \element[][12]CO by \cite{1999ApJ...512..761L}. The \element[][12]CO observations show a $3\degr \times 3\degr$ ring-shaped structure (see Figure \ref{fig:mbm16int}) with the strongest emission located at ($\alpha$:$\delta$:J2000.0) $\sim$ (3$^\mathrm{h}$24$^\mathrm{m}$:$12\degr 18\arcmin)$; ($\ell$:$b$) $\sim$ (171.1$^\circ$:$-35.9^\circ$). Before the work of Zucker et al., MBM 16 was considered one of the nearest HLMCs, at about 80 pc \citep[see][]{1988ApJ...327..356H}, half the distance of the Taurus complex. However, large-scale IRAS 100$\mu$m maps show that the Taurus dark clouds seem to have a significant dust extension down to $b\sim-45^\circ$ (see \cite {1988LNP...306..168M} and Figure 1 of \cite {2003ApJ...586.1111M}). This morphological connection and a similar velocity for MBM 16 and the Taurus dark clouds suggest that they are related. With the new distance, it seems plausible that it is part of a much larger complex, as are MBM 3 and MBM 40. Like those clouds, MBM 16 does not show any evidence of internal star formation.  Although \cite{2000A&A...356..157L} discovered four potential T Tauri candidates in the direction of MBM 16, two of the candidates are clearly in front of the cloud from their GAIA DR2 and EDR3 parallaxes (\cite{2016A&A...595A...2G}; \cite{2018A&A...616A...1G}; \cite{2021A&A...649A...1G} - \cite{2000A&A...356..157L} had assumed a distance to the cloud of 60-95 pc based on the earlier Hobbs et al. estimate), while the other (1RXS J032802.3+111441, a binary) is likely at a distance $\geq$ 190 pc, making its association with the cloud unclear.

The highest E(B-V) value in MBM 16 is 1.05 mag at ($\ell$:$b$) 
$\sim$ (171.2$^\circ$:$-37.3^\circ$) and several of the more intense dust concentrations or knots are near that value based on the \cite{1998ApJ...500..525S} database. The cloud's location embedded in the large dust structure south of the Taurus Dark Clouds implies that some of the reddening is associated with the southern extension of the latter. Nevertheless, the more intense E(B-V) knots in the cloud certainly have visual extinction greater than 1 mag, indicating the associated gas is translucent according to \cite {1988ApJ...334..771V}. Since the highest \element[][12]CO antenna temperatures in MBM 3 are 2-3 times greater than in MBM 16, the associated gas column densities may indicate a higher gas-to-dust ratio for MBM 3.

The \element[][12]CO map of MBM 16 has a coarse velocity resolution (0.65 km s$^{-1}$), limiting a dynamical analysis but it suffices to delineate the confines of the cloud. The majority of molecular gas lies between $\sim$ 6 km s$^{-1}$ and $\sim$ 9 km s$^{-1}$, with some exceptions at both lower and higher velocities (see \cite{2003ApJ...586.1111M}). The molecular gas is spatially concentrated in knots within the main structure.  Some lines of sight show a hint of asymmetry in the spectral lines, but no definitive conclusion can be drawn due to the low velocity resolution.  A molecular mass estimate based on the \element[][12]CO data is  320 $\pm$ 190 M$_\odot$ \citep{1997AAS...191.0713T}, making MBM 16 a few times more massive than MBM 3 and nearly an order of magnitude more than MBM 40.

\section{Archival data and observations}\label{sec:observations}
This section briefly describes the data we used for this work.

\subsection{CO mapping}
\cite{2006A&A...457..197S} describe in detail the \element[][12]CO (J$=$1-0) and \element[][13]CO (J$=$1-0) observations of MBM 3. The \element[][12]CO and \element[][13]CO were simultaneously observed in the spring of 2003 using the SEQUOIA focal plane array mounted on the Five College Radio Astronomy Observatory (FCRAO) 14-m radio telescope. The SEQUOIA array consisted of 16 dual-polarized pixels arranged in a 4$\times$4 pattern, which produced a 5\farcm9$\times$5\farcm9 footprint on the sky. The cloud was observed using the on-the-fly (OTF) mode with a single OFF position located at $\ell = 131 \degr$, $b = -45 \degr$ and was sampled at better than the Nyquist rate ($20 \arcsec$ spacing with beamsize at 115 GHz of $47 \arcsec$ and $49 \arcsec$ at 110 GHz). The velocity resolution per channel in the reduced data is  0.063 km s$^{-1}$ at 115 GHz and 0.066 km s$^{-1}$ at 110 GHz with the total bandwidth being 65 km s$^{-1}$ for \element[][12]CO and 68 km s$^{-1}$ for \element[][13]CO. The fidelity of the LSR velocity scale was checked by a new Onsala observation (see Appendix \ref{app:onsala_co}). The typical system temperature for the 2003 data was $\sim$800 K and after averaging the two polarizations, the typical rms noise was $\le$0.2 K per channel. We assume a beam filling factor of value one and the data are presented here in terms of antenna temperature, T$_A^*$; conversions to the main beam antenna temperature, T$_{mb}$, are made only when calculating the mass of the cloud (see Appendix \ref{app:mass}). The lower panel of Figure \ref{fig:secondary_cloud} shows averaged profiles of \element[][12]CO and \element[][13]CO over the whole cloud.

For MBM 16, the cloud was also sampled at better than the Nyquist rate (3.6$\arcmin$ spacing with an 8.4$\arcmin$ beam) in \element[][12]CO using the 1.2 m Harvard-Smithsonian CfA Millimeter Wave telescope \citep[] {1993ApJ...418..730D}.  The mapped region, extending in [$\ell, b$: 168$^\circ$ to 174$^\circ$, $-$34$^\circ$ to $-$40$^\circ$]\footnote {In 2000.0 ($\alpha, \delta$), the map is a rectangle with corners at (50.367$^\circ$, 15.434$^\circ$), (46.378$^\circ$,  10.860$^\circ$), (54.115$^\circ$, 12.044$^\circ$), and (49.832$^\circ$, 7.777$^\circ$).},  is shown by \cite{1999ApJ...512..761L} and, in more detail, by \cite {2003ApJ...586.1111M}. The velocity resolution was 0.65 km s$^{-1}$ and each of the 9409 spectra had rms values per channel of 0.1-0.3 K. MBM 16 is the large central structure in the original map (see Figure \ref{fig:mbm16int}  which shows only the central region of the original 6$^\circ \times$6$^\circ$ map) and its emission ranges from 5 to 11 km s$^{-1}$ in LSR velocity. Outside of the field shown in Figure \ref{fig:mbm16int} are unnamed molecular clouds to the northeast and southeast with LSR velocities in the 12-14 km s$^{-1}$ and $-$10 to $-$5 km s$^{-1}$ range, respectively, which may be associated with MBM 16, although we will not discuss them further in this paper.  \cite {2003ApJ...586.1111M} show position-velocity maps of the 6$^\circ \times$6$^\circ$ MBM 16 region, which show the three clouds as distinctly separated in velocity.

\subsection{Atomic hydrogen}

 We use the GALFA-\ion{H}{i} (Galactic Arecibo L-band Feed Array \ion{H}{i}) narrow band data archive \citep[see ][]{2011ApJS..194...20P,2018ApJS..234....2P} to study the atomic hydrogen in the two clouds and their surroundings.  This is an extended survey between $-1\degr \lesssim \delta \lesssim 38\degr$ with spatial resolution of about 4\arcmin \ and velocity resolution per channel of about 0.184 km s$^{-1}$ at 1.42 GHz. The GALFA-\ion{H}{i} data were obtained with the 305-m William E. Gordon radio telescope located in Arecibo, Puerto Rico.
The \ion{H}{i} data have different velocity and spatial resolution than the CO data, so to match them we performed velocity and spatial interpolations using the following procedure.

Firstly, we linearly resampled each \ion{H}{i} spectrum using the CO velocities as query points, bringing all spectra from different atomic and molecular species to the same velocity sampling. This procedure allowed us to more easily compare the line profiles and, because of the high S/N ratio, introduced no artifacts in the atomic hydrogen spectra.  Secondly, we then treated each \ion{H}{i} velocity slice as an image and performed a 2D linear interpolation using the CO grid as the array of query points. Since the \ion{H}{i} spectra change little over several beams, we are confident that this interpolation did not introduce any artifacts. However, there is an \ion{H}{i} artifact near MBM 3 and MBM 16 that is probably due to the 21 cm OTF mapping pattern (see Appendix \ref{app:artifact} for a discussion about the artifact near MBM 3. The conclusions are the same for MBM 16).

\subsection{MBM 3 Archival CH data}
\begin{figure}
  \centering
  \includegraphics[width=\hsize]{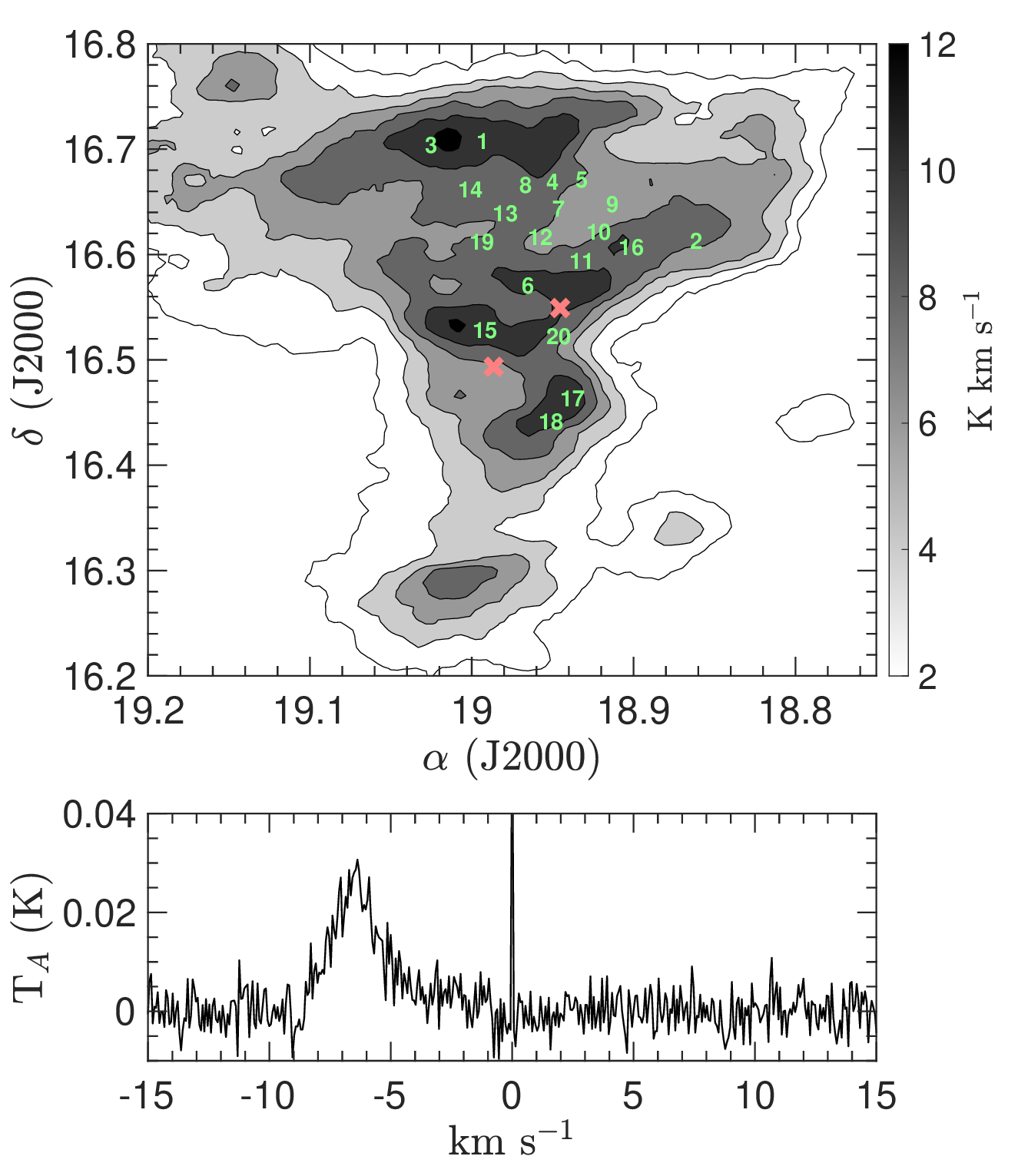}
    \caption{Positions of CH pointings. Numbers in green are the detections, and the red crosses are the non-detections. Gray-filled contours are the \element[][12]CO integrated antenna temperature. Below is shown the averaged CH spectrum using all the points where there is a detection. The spike at 0 km s$^{-1}$ is due to the autocorrelator and should be ignored.
    }
        \label{fig:ch_pointings}
\end{figure}

\cite{2010AJ....139..267C} describes the archival CH data for MBM 3 that we use in this work. The observations were performed using the Arecibo radio telescope during May 2003 and August 2004. The extreme weakness of the line precluded a more extensive map of the cloud in the time available, so only 22 sparse points were observed as reported in Figure \ref{fig:ch_pointings}, where the numbers in green indicate the detections, and the red crosses the non-detections. In general, the 3335 MHz CH line is extremely weak in this type of cloud, so to sample as many points as possible the integration time changed from one pointing to another, generally observing long enough to get about the same S/N {\it if} something was detected. If there was no hint of a line after 15-30 minutes, the integration was ended, even if the rms was greater than for lines of sight with detections. The data is censored by this procedure, which makes the statistical analysis more difficult.

For all points, the range of integration time was between 15 minutes to 90 minutes The brightness temperature (T$_B$) is obtained by dividing the antenna temperature (T$_A$) by the antenna efficiency, $\eta_B$, was $\sim$0.6 at 3.3 GHz for most observing conditions during the two observing runs. The typical brightness temperature of CH was a few tens of mK with unusually wide linewidths of $\sim$3 km s$^{-1}$. 
The beamwidth was $1 \farcm 3 \times 1 \farcm 6$ at 3335 MHz, the typical system noise temperatures were $\sim$30 K, and the rms noise was 10-20 mK. The velocity resolution was $0.068$ km s$^{-1}$ and the velocity range was 70 km s$^{-1}$. Given the stability of the digital electronics of the S-band receiver, only "ON" scans were taken to maximize the integration time on each point.  The bandpass profile was flattened using a polynomial fit.

The combined CH spectrum of the cloud is shown in the bottom panel of Figure \ref{fig:ch_pointings}  using all the pointings where there was a detection. The asymmetry of the line with a more pronounced redshifted wing may be a signature of the second cloud at $\sim$1 km s$^{-1}$, but the S/N is too low to make a definitive association.

\section{Discussion}\label{sec:discussion}

\subsection{Filamentation}\label{subsec:filamentation}
\cite{2024ApJ...968..131T} used the ALMA Compact Array to map a high-density star-forming region in the Corona Australis cloud at high spatial resolution, finding several filamentary structures with an average width of $\sim$ 10$^{-3}$ pc and a length of $\sim$ 10$^{-2}$ pc in C$^{18}$O and SO with column density of about 10$^{22}$ cm$^{-2}$.   \cite{2022MNRAS.514.3593T} and \cite{2023ApJ...948..109K} studied the 3D shape of the Musca filament, finding that observations are compatible with a sheet-like geometry that appears as a filament due to projection along the line of sight. Our previous study on MBM 40, our re-analysis of MBM 3, and MBM 16 archival data also point towards a combination of filaments and sheets in local low mass, diffuse,  molecular clouds.

\begin{figure*}
\centering
   \includegraphics[width=18cm]{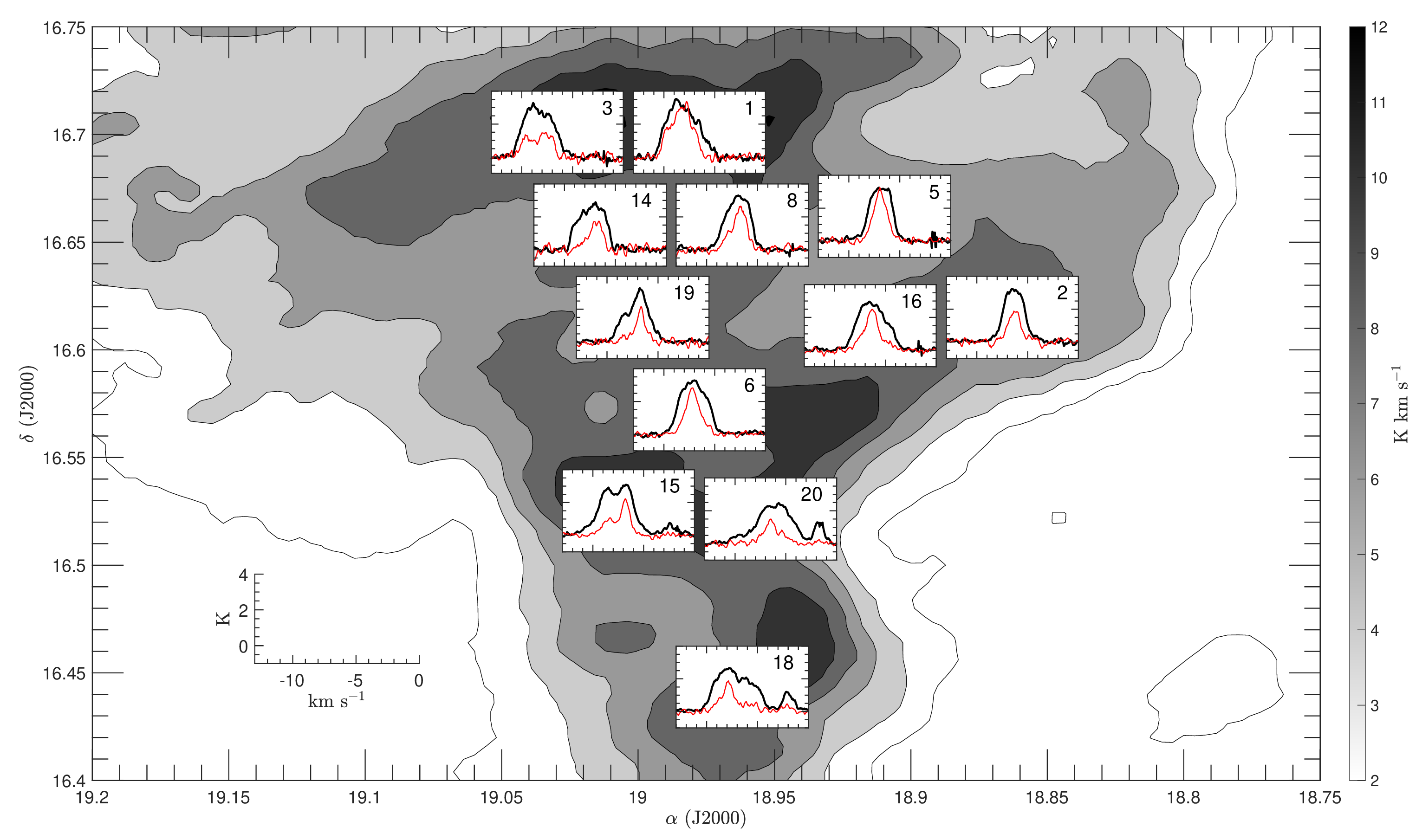}
     \caption{Selected positions of CO corresponding to the CH positions reported in Figure \ref{fig:ch_pointings}. The location of each subplot is indicative of pointings observed both in \element[][12]CO and \element[][13]CO. The filled contours are the same as the Figure \ref{fig:ch_pointings}. In each subplot, the black thick line is \element[][12]CO and the red thin line is $^{13}$CO multiplied by a factor of three for clarity. A frame with the same scale and range for each subplot is reported in the bottom-left corner of the figure.}
     \label{fig:CO_positions}
\end{figure*}

\begin{figure*}
\centering
   \includegraphics[width=18cm]{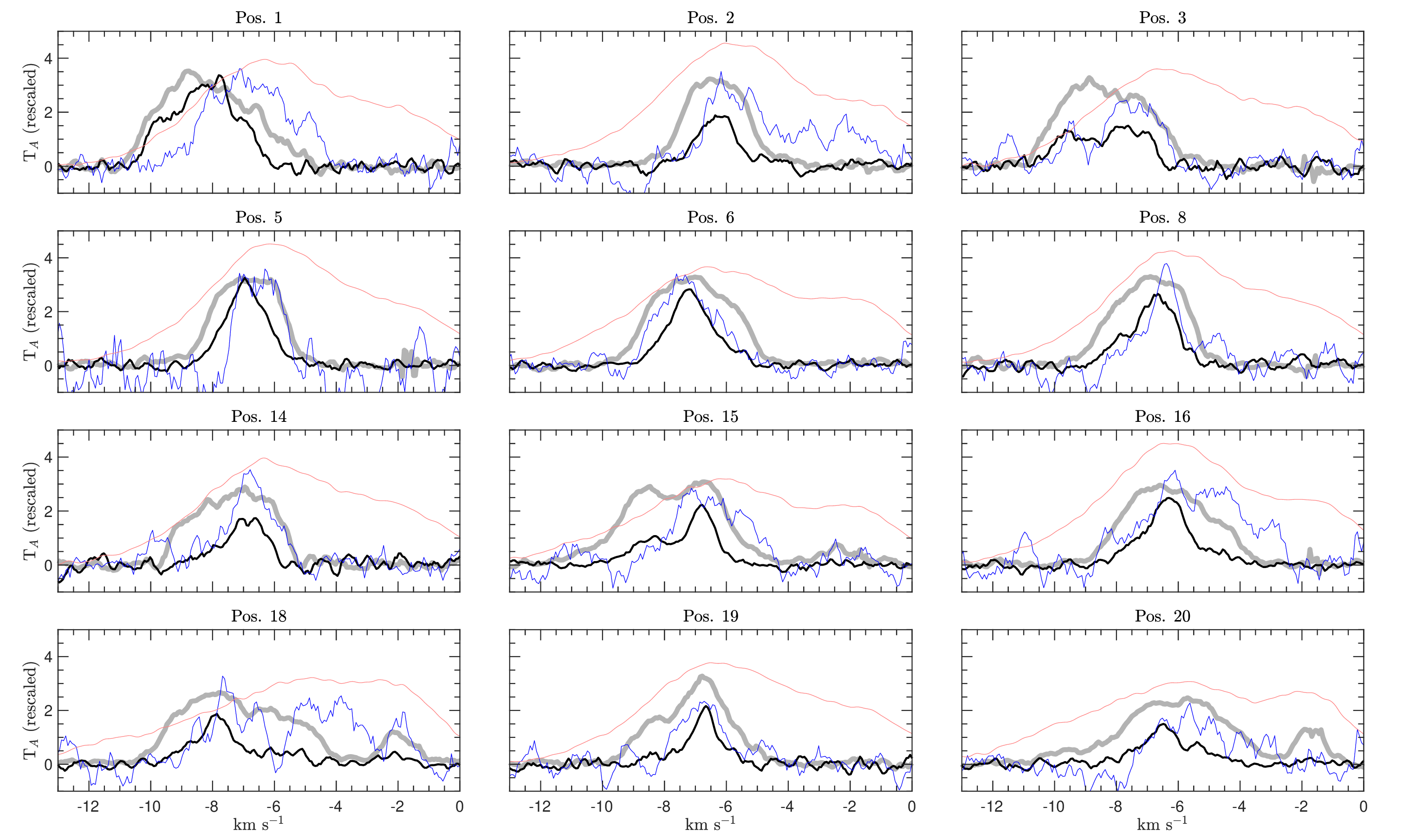}
     \caption{The grey thick line is \element[][12]CO, the black thick line is $^{13}$CO, the blue thin line is CH, and the red thin line is \ion{H}{i}. The \element[][12]CO profiles are reported without any rescaling, but the CH was multiplied by a factor of 60-130, depending on position, so it is possible to compare with \element[][12]CO. Similarly, the \ion{H}{i} was multiplied by a factor of 0.15 for all positions.}
     \label{fig:CO_CH_HI}
\end{figure*}

Averaging all spectra within the same cloud, as shown in the lower panel of Figure \ref{fig:secondary_cloud}, may be misleading. Different regions of the cloud show slightly different velocity centroids and different line shapes. Combining spectra from disconnected regions will produce a net profile whose linewidth is not directly linked with the local gas dynamics, but rather with the intensity-weighted distribution of the gas velocity within the whole cloud. The single line profiles, or an averaged profile within a small region (0.05 pc at most), show an asymmetric line profile which can result from non-thermal broadening and a superposition of separate flows. This appears to be the case for MBM 3, similar to what we found for MBM 40.

Figures \ref{fig:CO_positions} and \ref{fig:CO_CH_HI} shows spectra of \element[][12]CO, \element[][13]CO, CH and \ion{H}{i} at selected positions, and some positions show multiple components, even within the main cloud. For example, Positions 15, 18, and 20 show profiles from the outlier cloud and the main cloud formed by at least two overlapping filaments or sheets.  The linewidth, especially for  \element[][12]CO, is due to the superposition of two flows, one at $\sim$ 
$-$5 km s$^{-1}$ and the other at $\sim$ $-$9 km s$^{-1}$. Furthermore, a three component Gaussian decomposition of Position 15 yields very similar linewidths for each: $\sigma = 1 \pm 0.1$ km s$^{-1}$ for the blueshifted profile, $\sigma = 0.7 \pm 0.1$ km s$^{-1}$ for the redshifted profile and $\sigma = 0.8 \pm 0.1$ km s$^{-1}$ for the outlier cloud. A  similar result is obtained in Position 3, with two Gaussians linked with the main cloud, both with $\sigma \simeq 0.8$ km s$^{-1}$, as we found in MBM 40 \citep{2023A&A...676A.138M}. The blueshifted filament spans about 0.1$^{\circ}$ that, for a distance of $\sim$ 300 pc,  corresponds to a linear size of $\sim$0.6 pc.  Numerical simulations find similar properties \citep[see][]{2016MNRAS.457..375F}.

The \element[][13]CO profiles are systematically narrower than \element[][12]CO, but they show the same structure. For example, looking at Position 15, the double-peak structure in \element[][12]CO is also discernible in $^{13}$CO, but the peaks show different \element[][12]CO/\element[][13]CO ratio (hereafter $R_{CO}$), suggesting that the filaments have different optical depths. In Position 15, the filament traced by the redshifted peak shows $R_{CO} \sim 4 \pm 1$, and that traced by the blueshifted peak has $R_{CO} \sim 9 \pm 1$. Positions 18 and 20 are quite unusual because the optically thicker filament is blueshifted, oppositely compared with position 15, 19, and 14.

If the transverse optical depth within the same filament does not change too much, then the shift in velocity between Positions 15 and 18 can be purely topological: the twisting of the filament changes the velocity projected along the line of sight. If, instead, the gas optical depth does change along the filament, then the optically thinner peak in Position 15 is the optically thicker peak in Position 18. It is most likely that the observed profiles result from a combination of these two effects.  

Position 18, as well as position 2, shows a CH profile much wider than $^{12}$CO and $^{13}$CO, especially at lower velocities, i.e. $\sim$ $-$4 km s$^{-1}$. The low S/N does not allow a definitive statement, but the CH seems to be more pervasive where the CO is optically thin and, in general, where the gas is more diffuse. This view is compatible with what is observed in the \ion{H}{i} (see the lower panel in Figure \ref{fig:filament_HI}).

Using a number density tracer, such as the 4.8 GHz H$_2$CO line, may help to distinguish an optical depth effect from a pure topological twisting, since in the latter case the local gas density remains approximately constant along the filament. We are planning to observe H$_2$CO for the first time in different selected positions within MBM 3 in the near future to address this question.

As described above, MBM 16 presents an incomplete ring-shaped structure, with two different branches (see Figure \ref{fig:mbm16int}), separated in declination. Despite the poor velocity resolution, each filament is not monolithic in velocity, i.e. it is possible to see a systematic drift in both, consistent with kinked and tilted filaments. We will return to this in Section \ref{subsec:pvplotsMBM16}.

\begin{figure}
\centering
   \includegraphics[width=\hsize]{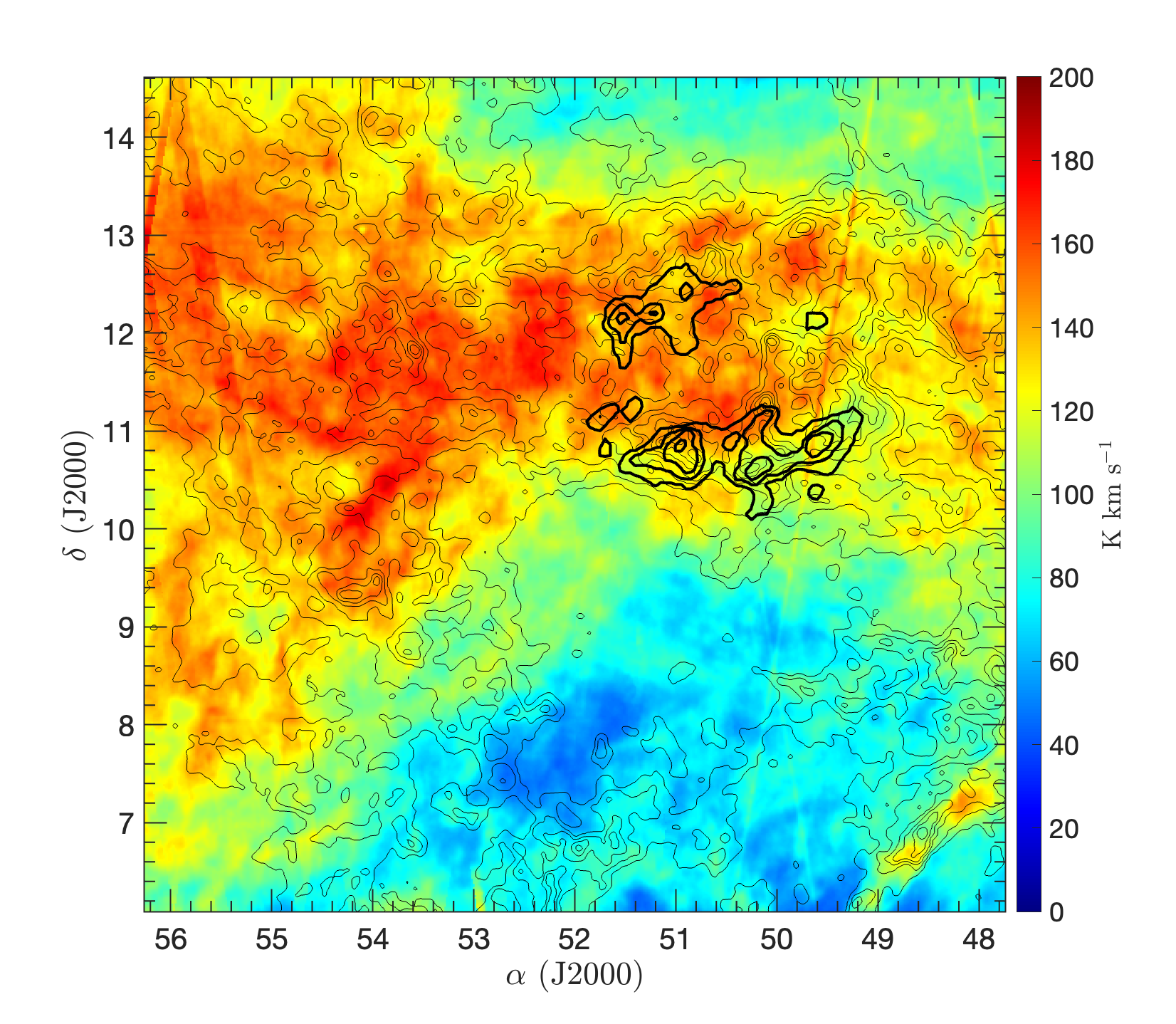}
     \caption{An ($8\degr \times 8\degr$) composite image of MBM16 and surroundings. The filled-colored image represents integrated \ion{H}{i} antenna temperature between 6 and 9 km s$^{-1}$, the thin contours depict the dust distribution as seen by 100 $\mu$m IRIS channel, and the thick contours are the integrated \element[][12]CO antenna temperature between the same interval of \ion{H}{i}. Note that MBM 16 lies amidst a wider atomic filament.}
     \label{fig:hi_filament_iris100}
\end{figure}

\subsection{Velocity slices in MBM 16}\label{subsec:vel_slices}

\begin{figure*}
\centering
   \includegraphics[width=18cm]{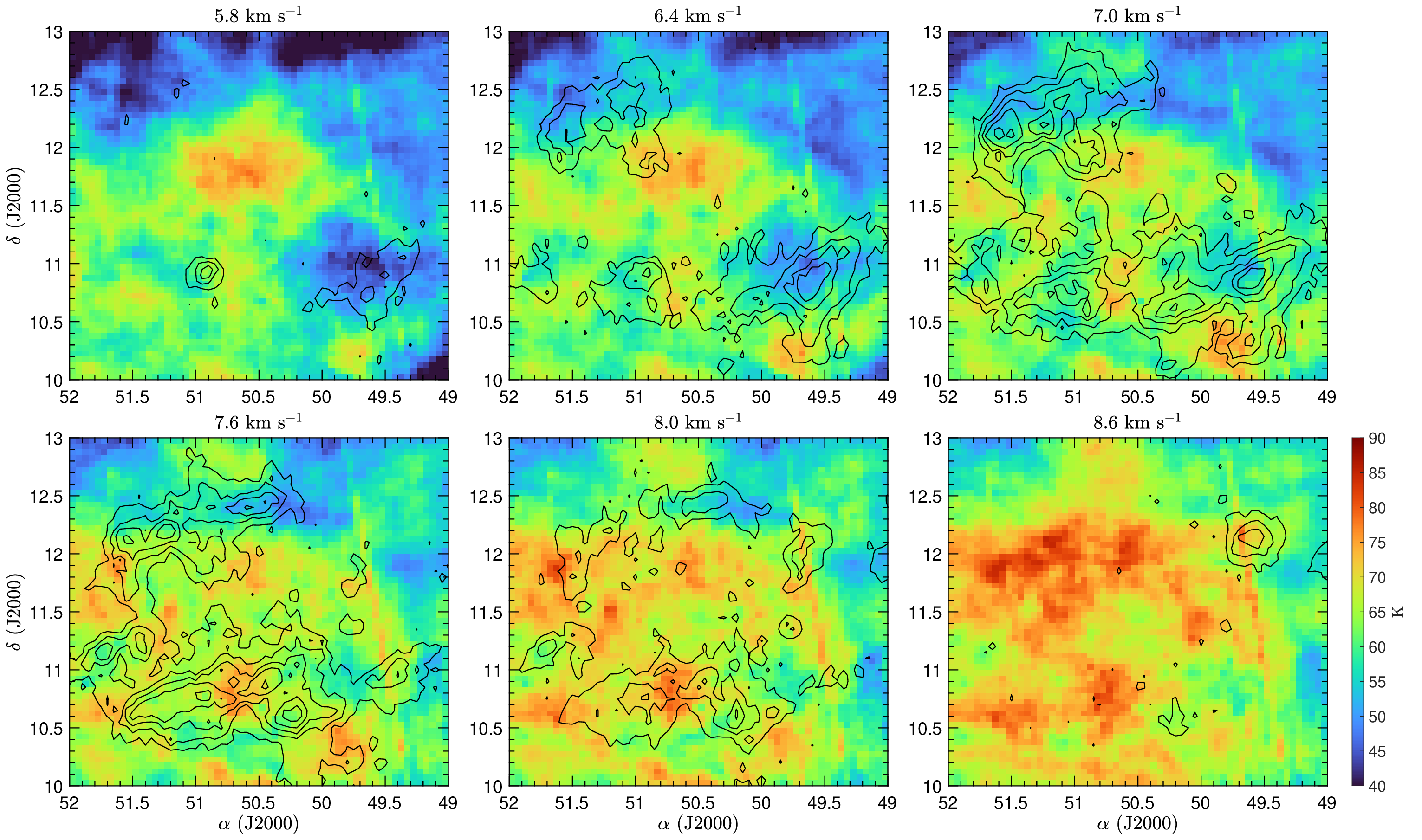}
     \caption{Velocity slices in MBM 16 discussed in Section \ref{subsec:vel_slices}. The contours show the \element[][12]CO distribution and the levels are from 0.5 K to 2.5 K in steps of 0.5 K, superposed on \ion{H}{i} (filled map). The colormap is the same in all panels, and the antenna temperature scale is shown in the lower-right corner. The tilted stripe on the right, particularly evident in the bottom panels, is an artifact caused by the \ion{H}{i} mapping technique (see Appendix \ref{app:artifact}).}
     \label{fig:vel_slices_mbm16}
\end{figure*}

Unlike MBM 40 \citep[see Figure 3 in][]{2022A&A...668L...9M}, the correlation between atomic and molecular gas is not as evident in the velocity integrated CO map of MBM 16 (see Figure \ref{fig:hi_filament_iris100}).
There is, however, a relationship between the atomic and molecular gas when examined in individual channels.
Figure \ref{fig:vel_slices_mbm16} shows the association between  \element[][12]CO and \ion{H}{i} in MBM 16. Each panel shows the same velocity slice of \ion{H}{i} (colored image) and of \element[][12]CO (black contours), in steps of 0.6 km s$^{-1}$.  Anticorrelations between the atomic and molecular gas are evident. For example, the 6.4 km s$^{-1}$ panel shows a molecular knot at $(\alpha, \delta) \sim (49.6\degr, 10.8\degr)$ overlying a \ion{H}{i}-poor region; the southern filament also presents molecular knots in the 7.6 and 8.0 km s$^{-1}$ panels that spatially alternate with \ion{H}{i} knots (the lack of atomic gas at $\alpha \sim 51.1\degr$ and $\alpha \sim 50.2\degr$, where the \element[][12]CO shows a higher antenna temperature).  This anti-correlation can be explained as a decrease of atomic hydrogen in favor of molecular gas, indicative of a phase transition, as we previously found in MBM 40.  This seems to be a common property in HLMCs
\citep{1994ApJ...434..162G}.

\subsection{Turbulence and spatial scale of intermittency}
Translucent and diffuse clouds are short-lived ($\sim$10$^6$ yrs - \citealt[]{2005ApJ...618..344V}; \citealt{1985ApJ...295..402M}), forming by a phase transition from the atomic phase. These clouds are often parts of larger atomic structures, i.e., \ion{H}{i} filaments.  For MBM 40 \citep[see][]{2023A&A...676A.138M} we found evidence of an \ion{H}{i} structure that envelops the molecular cloud and drives the internal dynamics by injecting kinetic energy through shear flows that fuel turbulent motions.

The turbulent nature of the internal dynamics of molecular clouds produces several observational hints in MBM 3. Considering only the thermal broadening, the \element[][12]CO linewidth at 20 - 30 K would be about 0.1 km s$^{-1}$, but the observed lines are systematically broader.  Furthermore, the profiles are skewed in velocity, e.g., see the bottom panel of Figure \ref{fig:secondary_cloud}. This asymmetry can be produced by a superposition of several structures, each of which has internal superthermal velocities compatible with unresolved turbulence.  Thus, if we assume the trend is tracing individual filaments with different densities, the velocity trends along the filaments imply a geometric twist.

Turbulence is intrinsically coherent on small scales and isotropic on large scales.  It is characterized by vortices, filaments, and twisting that possess a high level of correlation measurable by a variety of methods.  For example, the relaxation of the Probability Distribution Function (PDF) of velocity centroids with spatial lag, or the analysis of Structure Functions (SF), that characterize the spatial correlation of observational proxies, such as velocity centroids or line intensities \citep[see][]{1985ApJ...295..466K, 1985ApJ...295..479D, 1991JFM...225....1V, 2000tufl.book.....P, 2003ApJ...583..308P}. In general, these tools provide a characteristic coherence scale of the flow: for  PDFs, it is the lag at which the distribution is completely relaxed, while for SF, it is the lag at which the SF is no longer a power law (or changes exponent).

\begin{figure}
  \centering
  \includegraphics[width=\hsize]{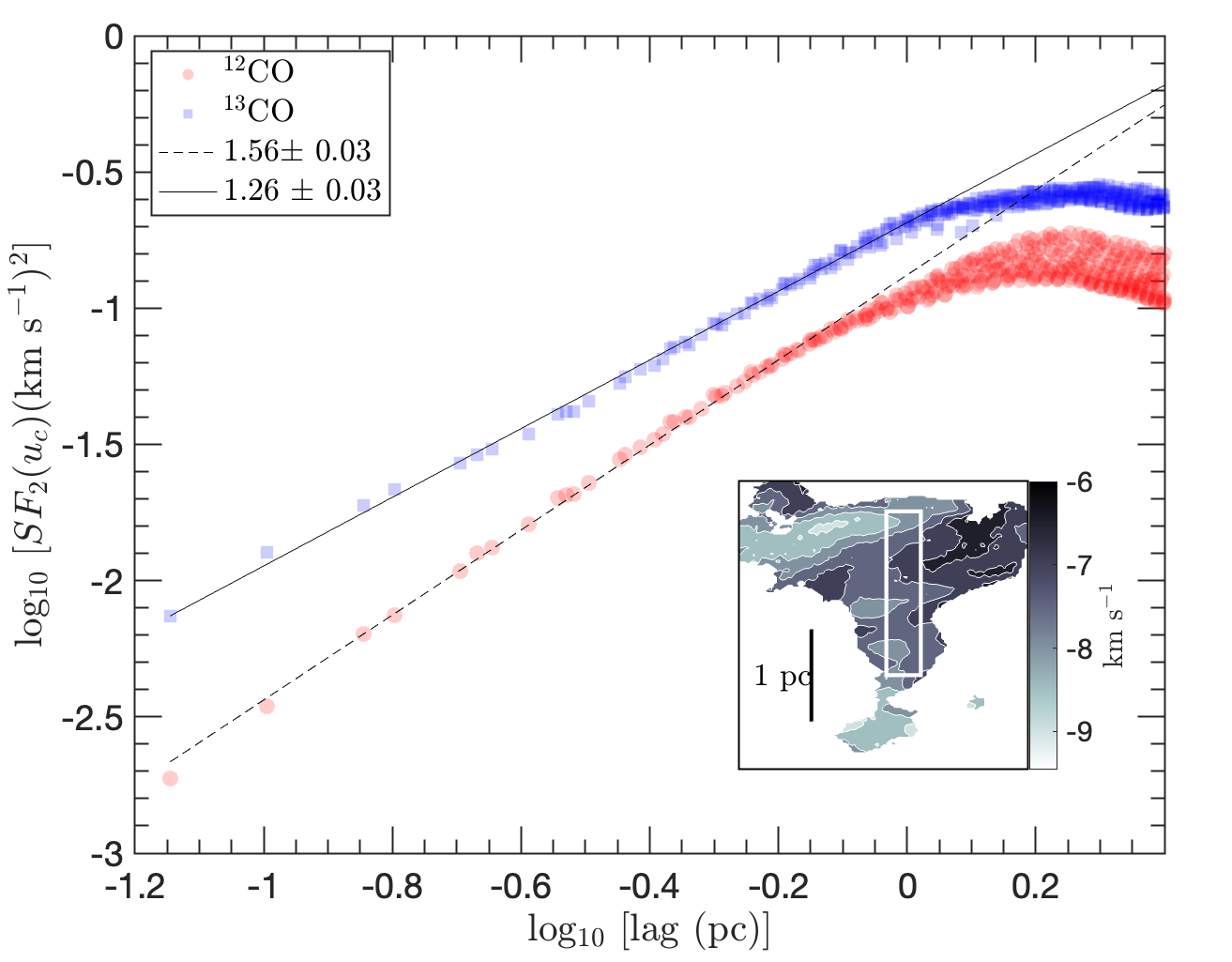}
    \caption{Second Structure Function ({\rm SF$_2$}) of velocity centroids (both \element[][12]CO and \element[][13]CO). The slopes of the best fits are reported in the legend. \element[][13]CO shows a scaling law only for a portion of the lags. The insert shows the \element[][12]CO centroid map, and the region used for the SF computation is within the white rectangle.}
        \label{fig:SF2_v}
\end{figure}

The SF for a map of some quantity A of order $p$ is defined as
\begin{equation}\label{eq:SF}
    {\rm SF_p} (A,\mathbf{r}) = \langle \left| A(\mathbf{r}) - A(\mathbf{r}+\delta \mathbf{r}) \right|^p \rangle ,
\end{equation}
where the angle brackets indicate a spatial mean over the same lag $\mathbf{\delta r}$. Figure \ref{fig:SF2_v} shows the SF of order $p=2$, ${\rm SF_2}(\delta r)$, for the velocity centroids. For small lags, the SF is well approximated by a power law for both \element[][12]CO and \element[][13]CO, with the exponents indicated directly in the legend. For larger lags, i.e. $\delta \mathbf{r} > 1$ pc, the SF is no longer a power law, and we take 1 pc as the structural scale for this cloud, of the same order as we found for MBM 40 and compatible with a turbulent flow.

The $^{12}$CO and $^{13}$CO structure functions  show slightly different power laws within the same sample, even though the two species are well mixed. $^{13}$CO is a rarer isotope; therefore, it traces the denser gas within the cloud. On the other hand, the $^{12}$CO also picks up the more diffuse gas that surrounds the core of the molecular cloud. In this sense, the SF of $^{12}$CO traces a combination of two different gas regimes, whereas the $^{13}$CO does not. The scale at which the structure function is no longer a power law is also slightly different between the two species. This indicates that the turbulent regime is different between the outlier gas and the core, compatible with an energy injection on a bigger scale, as previously found in MBM 40. It is also worth mentioning that the scaling laws are not compatible with an incompressible, isotropic Kolmogorov-type turbulence, where SF$_2$ $\propto \ell^{2/3}$ is predicted \citep[][]{1941DoSSR..30..301K,1941DoSSR..32...16K}. This may indicate that the turbulence is neither incompressible nor isotropic and that what appears to be a monolithic structure is also a superposition of filaments with different turbulent characteristics.

Intermittency is probably the strongest evidence of turbulent motion, seen as large deviations in the velocity of rare events that occur with a systematically higher frequency than in an uncorrelated process \citep[][]{1990ApJ...359..344F,2009A&A...507..355F}. It can be traced by large deviations from the Gaussian distribution of the PDF's wings or by looking at the wings of a single line of sight profile with a good S/N (noting this method can be used only if the line of sight is optically thin and the spectrum is tracing all the gas, see \cite{2023A&A...676A.138M}. For MBM 3, we do not have data with sufficiently high S/N to detect these variations in either the PDF or in the available single line profiles.\footnote{ As a speculation, we note that if the turbulence spectrum is multifractal,  the high latitude clouds may actually represent the intermittency of the larger scale turbulence that, in turn, have substructures organized by the shear flows at the source scale provoked by different dissipation processes \citep[e.g.][]{2019JFM...867P...1D}.}

\subsection{Relationship with \ion{H}{i} in MBM 3}\label{subsec:hi}

As Figure \ref{fig:optic_hi_co} shows, the relationship between molecular and atomic hydrogen is not as clear in MBM 3 as we found in MBM 40 \citep[see][]{2022A&A...668L...9M}, where the \ion{H}{i} in the appropriate velocity range is shaped in a cocoon-structure that embraces the inner molecular cloud.  The \ion{H}{i} profiles in  MBM 3  show at least three components, like those for MBM 40. One is rather diffuse, spanning between $-$20 km s$^{-1}$ to 10 km s$^{-1}$ that contributes to the wings of the profiles and whose origin is unknown \citep[see, e.g.,][]{1994AJ....107..287V}. This extremely diffuse component changes within the cloud, since in the range of [$-$15, $-$10] km s$^{-1}$ is stronger in the stem relative to the anvil. In contrast, the redshifted wing appears to be virtually identical in both regions. It is not clear if the excess in the blueshifted wing is associated with the molecular cloud or if it is just due to line of sight atomic gas. Besides this diffuse component, the profiles generally show a two-peaked shape: the peak at $\sim -6$ km s$^{-1}$ is associated with the main cloud, and the peak at $\sim -3$ km s$^{-1}$ with the outlier cloud. Panels in Figure \ref{fig:CO_CH_HI} report the \ion{H}{i} profiles depicted in red.  Although the profiles change from one position to another, the global shape is nearly the same, with two enhancements at $\sim -6$km s$^{-1}$ and at $\sim -3$km s$^{-1}$ over the diffuse component.

\begin{figure}
  \centering
  \includegraphics[width=\hsize]{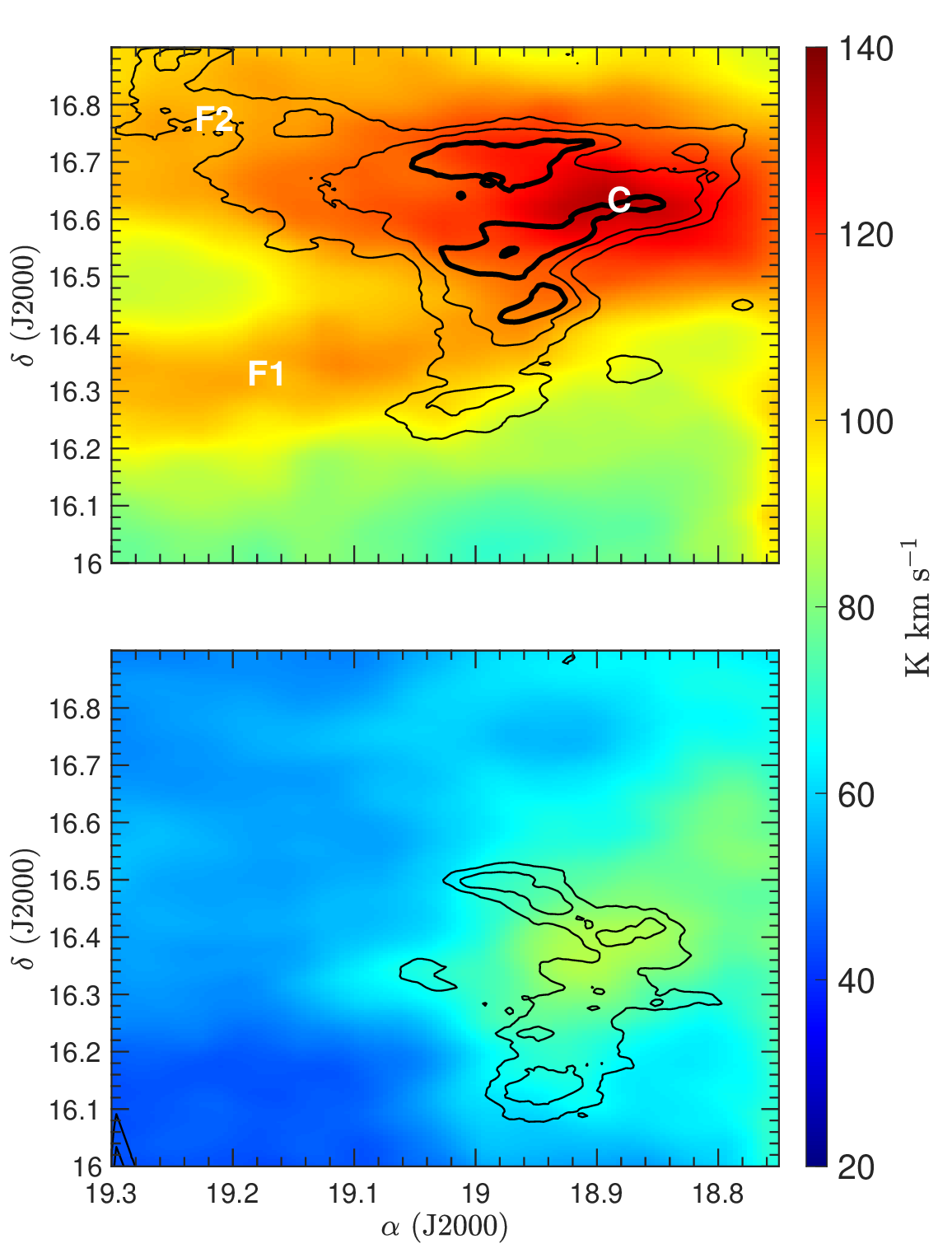}
    \caption{\ion{H}{i} integrated antenna temperature for the primary cloud (top panel, integrated from 
    $-$10 to $-$4 km s$^{-1}$) and for the outlier cloud (bottom panel, integrated from $-$3.5 to 1 km s$^{-1}$). The contours in each panel are from the \element[][12]CO, integrated on the same velocity range of \ion{H}{i} for each cloud. The primary cloud contours are 3, 6, and 9 (thick contour) K km s$^{-1}$, and the outlier cloud contours are 1.5 and 2.5 K km s$^{-1}$. The \ion{H}{i} colorbar is the same in both panels. Hence, the atomic hydrogen linked with the outlier cloud is distinctly less abundant than the hydrogen associated with the primary cloud. In the top panel are reported the positions of the selected \ion{H}{i} profiles of Figure \ref{fig:filament_HI}. See Section \ref{subsec:hi} for further details.}
        \label{fig:hi_pri_sec}
\end{figure}

\begin{figure}
  \centering
  \includegraphics[width=\hsize]{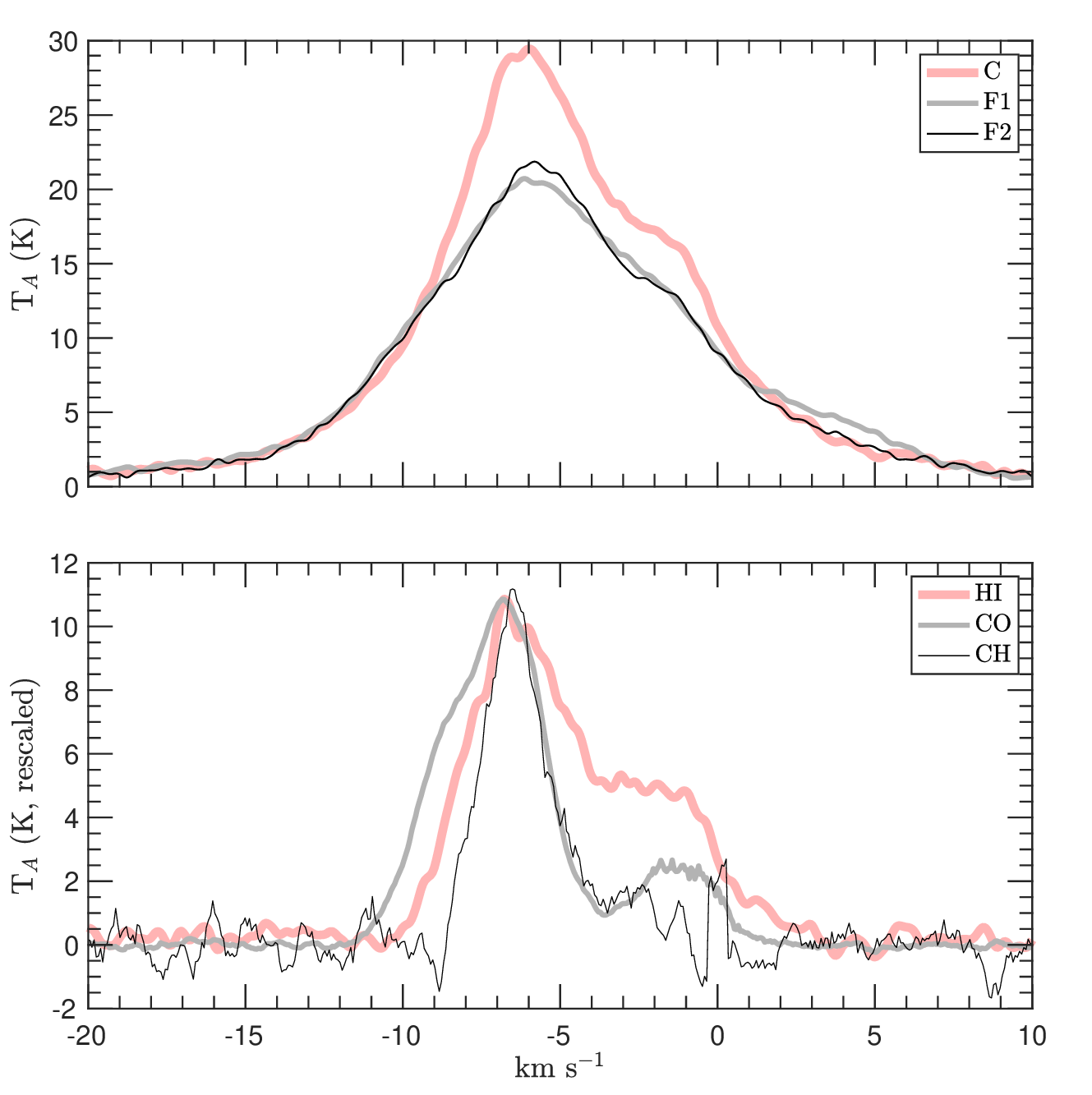}
    \caption{(Top panel) \ion{H}{i} profiles for the selected positions labeled in Figure \ref{fig:hi_pri_sec}. F1 and F2 are located inside the candidate filaments, and the profiles are surprisingly similar, whereas the C profile is linked with the MBM 3 core. (Bottom panel) The pale red thick line is the result of subtraction between C and F1, the grey thick line is the \element[][12]CO emission averaged over the whole cloud and the thin black line is CH. Note that the \element[][12]CO and CH are scaled to \ion{H}{i} to better compare the line shape of atomic and molecular gas.}
        \label{fig:filament_HI}
\end{figure}

Integrating over the whole velocity range produces an ambiguous result since the spatial correlation is masked by the contribution of diffuse gas. This is more dramatic for MBM 3 than we found in MBM 40. Figure \ref{fig:optic_hi_co} shows that most of the atomic emission is roughly aligned with the core of the molecular gas, but the correlation is poor. If we select a narrower velocity range for the integration, the atomic gas shows a greater spatial correlation. Figure \ref{fig:hi_pri_sec} shows the results for both the primary cloud and the outlier. In the upper panel, the velocity range is the same as the \element[][12]CO linewidth of the main cloud, [$-$10, $-$4] km s$^{-1}$, and although the spatial correlation is still poor, the bulk of the atomic gas is located in the anvil, where also the \element[][12]CO emission is quite strong. The lower panel shows the same result for the outlier cloud, using [$-$3.5,1] km s$^{-1}$.  Although the spatial correlation is still poor,  the enhancement of atomic gas roughly matches the \element[][12]CO emission.

We decomposed each \ion{H}{i} profile \citep[see, for example,][]{2015AJ....149..138L, 2021ApJS..256...37M} and then matched one (or more) component(s) with the corresponding molecular feature.  We find that the molecular gas is not completely linked with any single component or with the diffuse gas at higher velocities. This is evident in Figure \ref{fig:CO_CH_HI}: the molecular velocity peaks are, in most cases, aligned with the atomic velocity peaks but the \ion{H}{i} lines are much broader than those of the molecular gas, probably due to blending of diffuse gas that enhances the wing velocity channels and which is not directly associated with the cloud.

To investigate this further, we selected two positions away from MBM 3 (F1 and F2 in the top panel of Figure \ref{fig:hi_pri_sec}) and one inside the anvil (position C) and compared the profiles. The result is shown in Figure \ref{fig:filament_HI}: the off positions are virtually identical and the C position shows a double-peaked profile between $-$10 and 0 km s$^{-1}$. The bottom panel shows the cloud \ion{H}{i} profile once the F1 emission is removed (depicted in light red). After this procedure, the \ion{H}{i} profiles are more similar to the molecular profiles:  the linewidth now agrees well with the average linewidth of the molecular gas. The bottom panel also shows the averaged profiles of \element[][12]CO and CH.

The F1 position appears to be located inside a filament that spans the entire field, as shown in the 
top panel of Figure \ref{fig:hi_pri_sec}.  This is supported by the similarity between single profiles within the filament. Furthermore, the F2 profile is similar to those within the filament, so this is consistent with an arc-shaped structure that extends from the southeastern corner to the northwestern corner, and the molecular cloud then sits inside this filament.  After subtracting the filament, the \ion{H}{i} and molecular spectra are comparable. Both \ion{H}{i} and \element[][12]CO show two peaks, but their ratios are different. In the molecular gas, the ratio between the main cloud and outlier cloud peaks is about 5, while the atomic gas ratio is about 2; this may indicate that the conversion from atomic to molecular gas is more efficient in the main cloud than in the outlier.

\subsection{MBM 3 Position-Velocity plots}\label{subsec:pvplotsMBM3}
\begin{figure}
  \centering
  \includegraphics[width=\hsize]{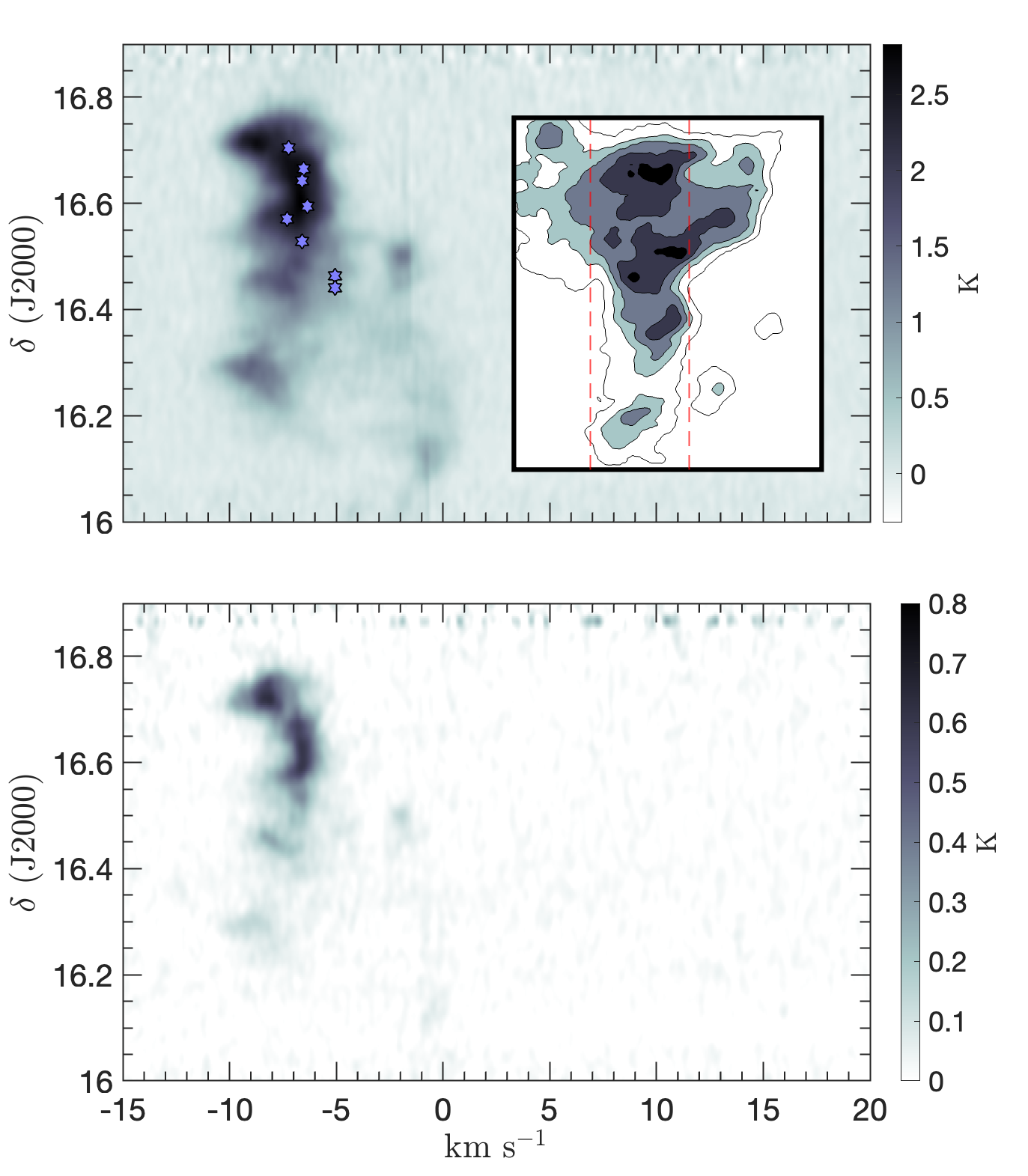}
    \caption{Position-velocity plot of \element[][12]CO (top panel) and \element[][13]CO (bottom panel) mean antenna temperature along declination for MBM 3. Blue stars indicate the CH velocity centroids. In the insertion, we show the RA range (red dashed lines) within which the PV plot is computed.}
        \label{fig:pv_plot_co}
\end{figure}

\begin{figure}
  \centering
  \includegraphics[width=\hsize]{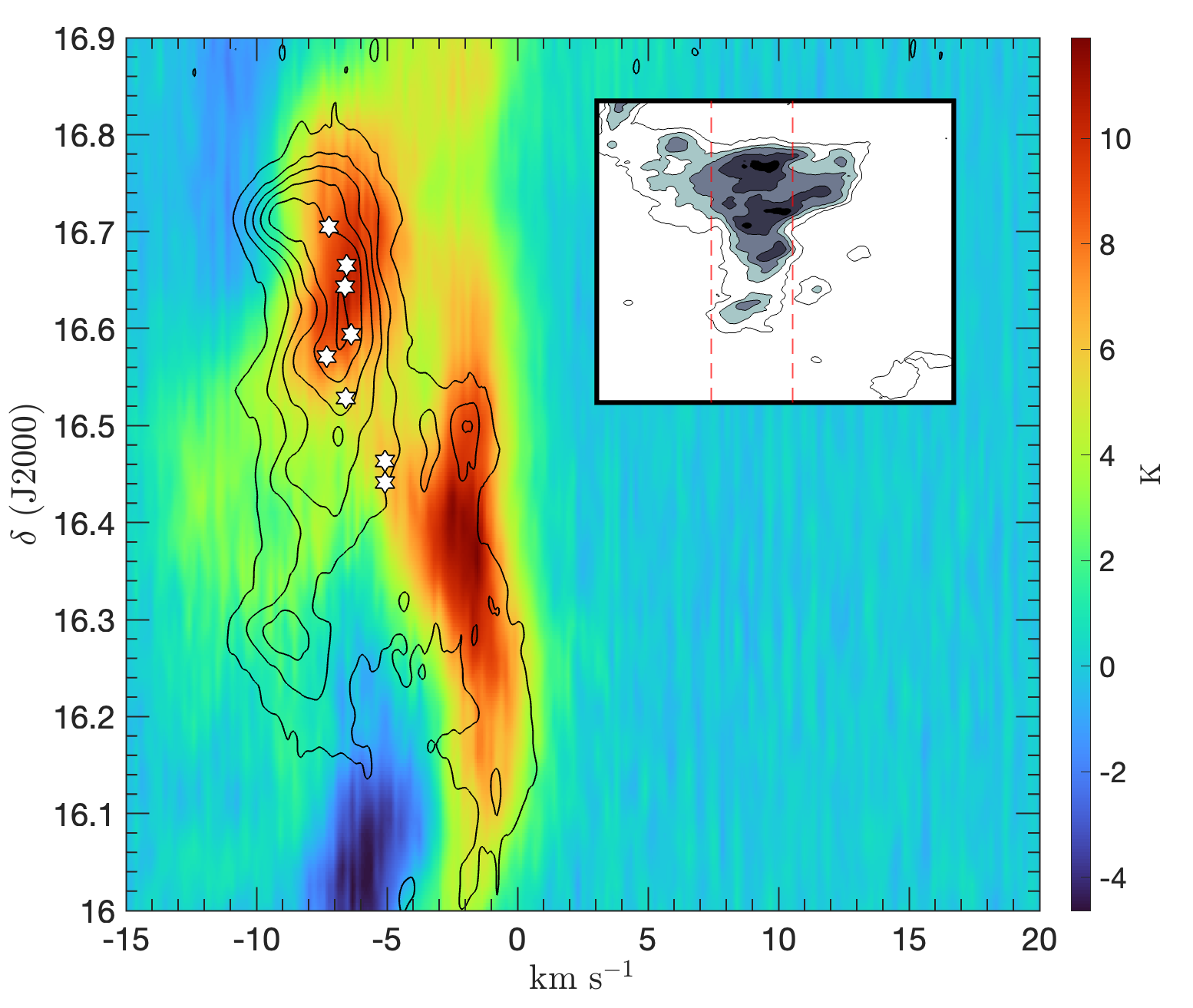}
    \caption{Position-Velocity plot of \ion{H}{i} (in color) with diffuse component subtracted (see text). The black contours show \element[][12]CO mean antenna temperature between 0.2 K and 3 K in steps of 0.5 K. White stars indicate the CH velocity centroids. In the insert, we show the RA range (red dashed lines) within which the PV plot is computed. Negative \ion{H}{i} antenna temperature arises from the diffuse component subtraction and should be ignored. For a description of extremely diffuse component subtraction, see \cite{2022A&A...668L...9M}.}
        \label{fig:HI_PV_plot}
\end{figure}

Figure \ref{fig:pv_plot_co} shows the Position-Velocity (PV) plot  for \element[][12]CO (upper panel) and $^{13}$CO lower panel), computed along the declination within the stem.  The cloud is rather compact in velocity, especially in \element[][13]CO, with the strongest emission on the cloud's northern side. Below $\delta =$ 16.5$\degr$ the \element[][12]CO shows a broader emission, with the velocity spanning the interval from $-$10 to 0 km s$^{-1}$. This emission is due to the superposition of the two clouds, which have different bulk velocities, and it is also visible in the \element[][13]CO, although systematically fainter. The CH centroid velocities are also displayed in the top panel and show a velocity drift moving southward through the stem. Figure \ref{fig:HI_PV_plot} displays a combination of atomic and molecular gas and highlights the relation between them. The \ion{H}{i} emission, as expected, is much broader both in position and in velocity, pointing out that the molecular gas is generally condensing from the atomic hydrogen.

The \ion{H}{i} PV plot shows two main structures, the northern one of the main cloud and the southern one that is the outlier cloud. Since the \ion{H}{i} and CO do not show any gap in declination, it is possible that the two clouds are actually parts of the same structure, with the molecular gas embedded in the atomic component. This inference is reinforced by the two lower CH observations, which show velocities midway between the two clouds. 

Both \ion{H}{i} and molecular gas show a shift in velocity of about 6 km s$^{-1}$ over 0.6$\degr$ in declination corresponding to 1.9 km s$^{-1}$ pc$^{-1}$ for a distance of 300 pc, that is similar to what we found for MBM 40.  While this may be evidence for a large-scale shear flow that runs along the stem, considering the topological interpretation we advocated in our previous paper, a combination of projection and dynamical effects is likely the explanation for this gradient.

\subsection{MBM 16 Position-Velocity plots}\label{subsec:pvplotsMBM16}

\begin{figure}
\centering
   \includegraphics[width=\hsize]{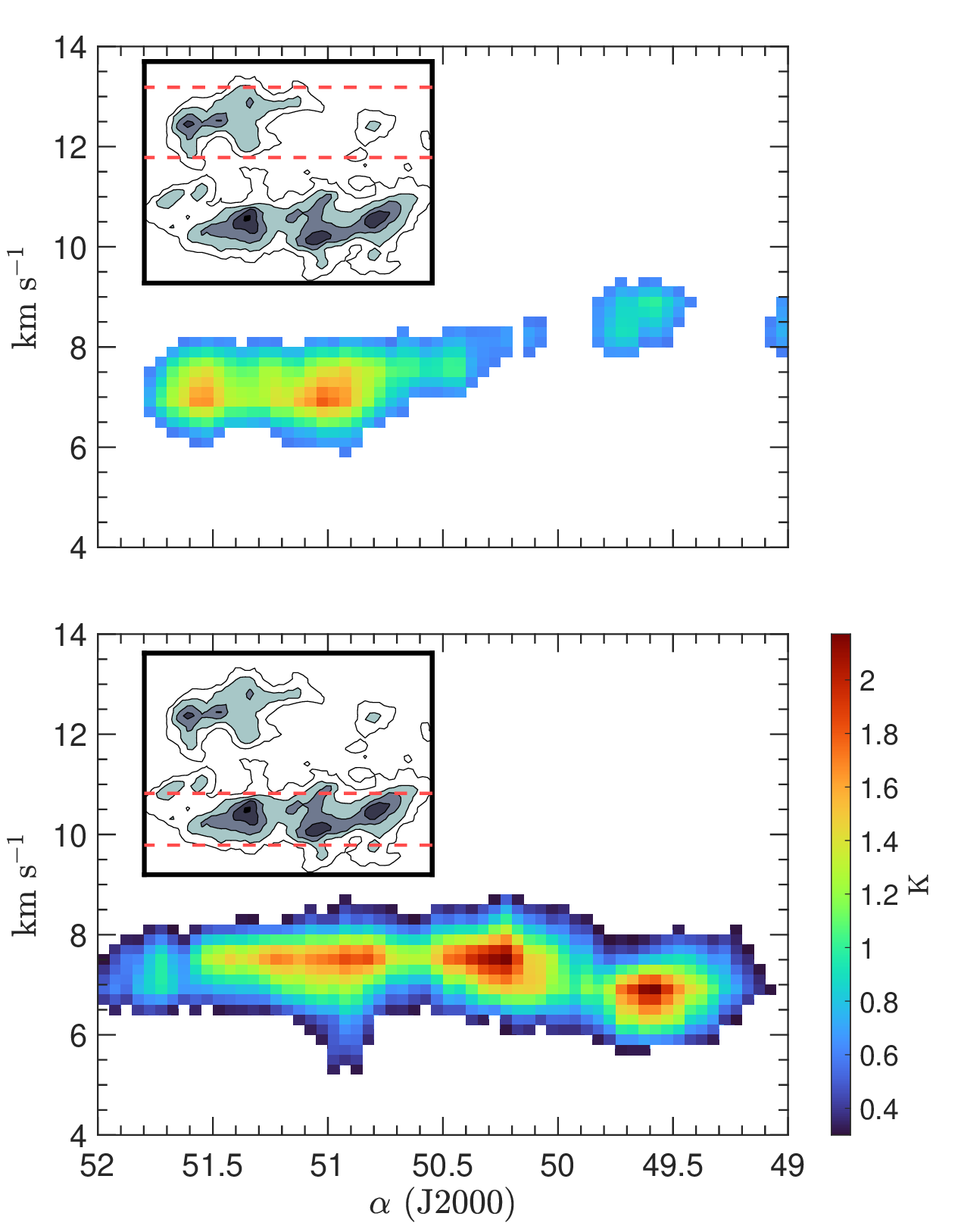}
     \caption{ Position-Velocity plots for MBM 16. The top and bottom panels show, respectively, the velocity distribution along the right ascension axis for the northern and southern portions of MBM 16. The PV plots are performed by averaging the spectra along declination only within the red dotted lines in each insert. The colorbar of the averaged antenna temperature in the lower right corner is the same for both panels.}
     \label{fig:pvMBM16}
\end{figure}

\begin{figure}
\centering
   \includegraphics[angle=270, width=\hsize]{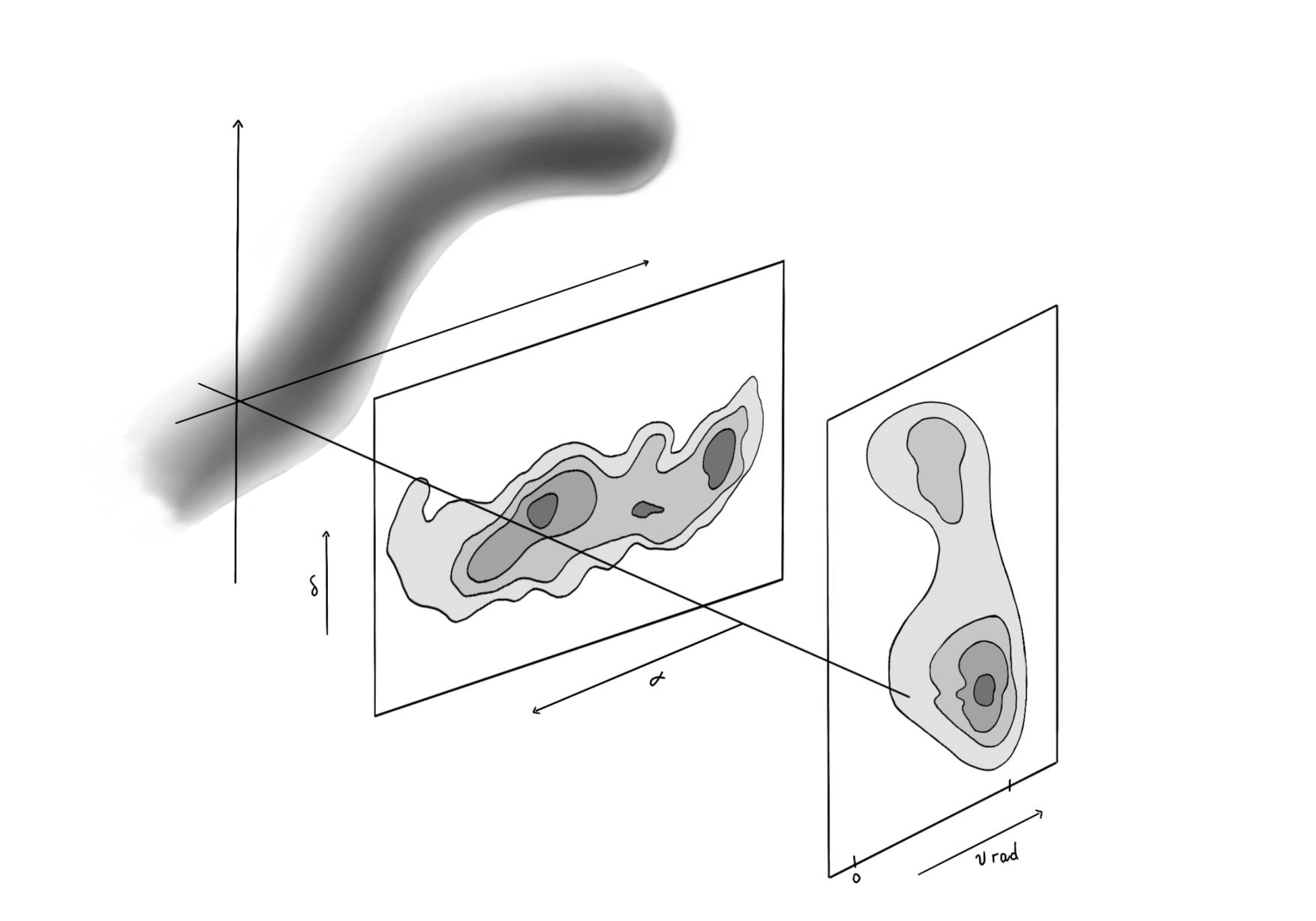}
     \caption{A schematic cartoon of how the MBM 16 position-velocity plots shown in Figure \ref{fig:pvMBM16} can result from projection effects of a writhing filament, similar to MBM 3 (Figures \ref{fig:pv_plot_co}, \ref{fig:HI_PV_plot}) and MBM 40 \citep[]{2023A&A...676A.138M}.  The vertical axis in the PV cartoon is the right ascension.}
     \label{fig:cartoon}
\end{figure}

Figure \ref{fig:pvMBM16} shows the right ascension  PV plots of MBM 16. The upper panel shows the cut, averaged in declination, over the northern branch of the cloud. The insert with the $^{12}$CO filled contours highlights with two dashed lines the region within which the average in declination was performed. The lower panel shows the declination summed PV plot for the southern branch. The velocity scale,   the right ascension interval, and the averaged antenna temperature scale are the same for both panels. The filaments show almost the same velocity width of about 2 km s$^{-1}$ and a drift in velocity for $\alpha \lesssim 52.2\degr$. The southern filament shows three different knots, also visible in the integrated antenna temperature in the insert, almost equally spaced (about 0.6\degr, which corresponds to $\sim 1.8$ pc for a distance of 170 pc). The eastern knots on the northern filament are separated by almost the same distance. This could indicate coherent structure. Considering the knots at $\sim 51\degr$ and at $\sim 49.6\degr$ (which correspond to a physical distance of $\sim 4.2$ pc), the velocity difference is about 1.6 km $^{-1}$; therefore, the velocity drift between these two knots is about 0.4 km s$^{-1}$ pc$^{-1}$. For the southern filament, considering the knots at $49.6\degr$ and at $50.2\degr$ (physical distance of 1.8 pc), with a velocity difference of 0.7 km s$^{-1}$,  the velocity drift $\sim 0.4$ km s$^{-1}$ pc$^{-1}$,  comparable to the northern filament. We previously found \citep[see][]{2023A&A...676A.138M} a similar behaviour in MBM 40, although the velocity drift along its western filament was more pronounced (i.e, about 1 km s$^{-1}$ pc$^{-1}$), indicating that shears in velocity are common in HLMCs and can shape their structure.  A schematic illustrating our interpretation of the filamentation traced by the PV plot is shown in Figure \ref{fig:cartoon}.

\begin{figure}
\centering
   \includegraphics[width=\hsize]{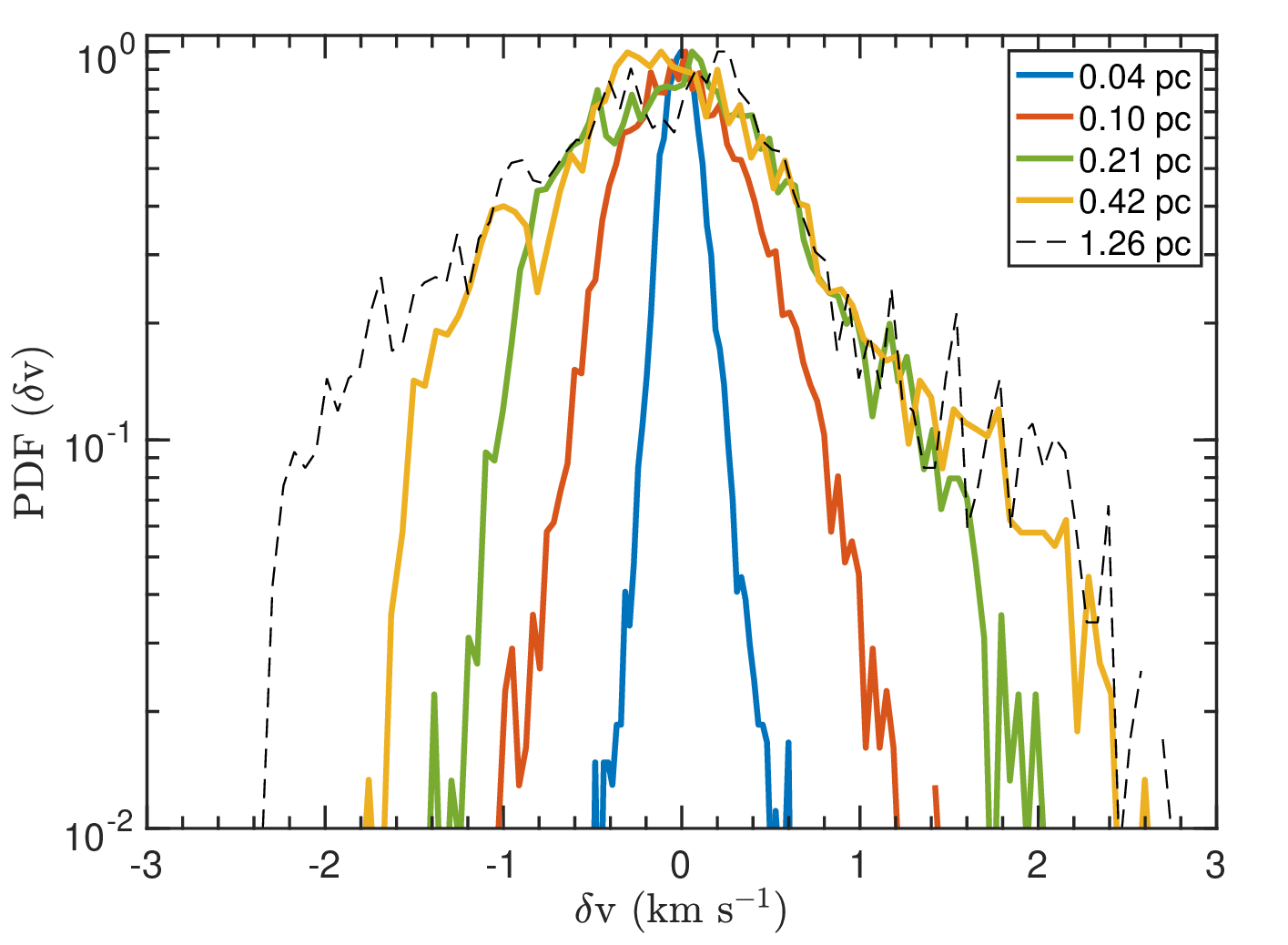}
     \caption{ Probability Distribution Function (PDF) of velocity centroids \citep[see][]{1985ApJ...295..466K} evaluated at different spatial lags, only along the declination axis within the stem (see the insert in Figure \ref{fig:SF2_v}).}
        \label{fig:PDF_rel}
\end{figure}

The velocity centroid Probability Distribution Function (PDF) for  MBM 3 is shown in Figure \ref{fig:PDF_rel}.  The PDF was evaluated along the declination axis within the stem for lags of 2, 5, 10, 20, 60 beams, which correspond to 0.04, 0.10, 0.21, 0.42, 1.26 pc for an assumed distance of 300 pc. The PDF is underdispersed for small lags ($\leq 5$, 0.1 pc) and it is symmetric to $\delta v = 0$; this is in line with the picture of a turbulent medium, since it maintains more coherence in flow properties with respect to an uncorrelated process. At greater lags, the PDF relaxes to a broader asymmetric distribution, indicating that the turbulent structure of the cloud is not spatially isotropic. Looking at the maximum lag measured of 1.26 pc (dashed black line), the red wing of the PDF is very similar to the 0.42 pc lag. This suggests that the correlation scale is about 0.2 to 0.4 pc,  similar to our previous result in MBM 40, for which we found $\sim$ 0.4 pc. The difference between the blue wing of the PDFs could indicate a superposition of multiple flows with slightly different central velocities and different spatial correlation scales. The correlation scale estimated with the PDF is compatible with that obtained analysing the second Structure Function (see Figure \ref{fig:SF2_v}, where the scale at which the SF is no longer a power law is about 0.6 - 1 pc.)

\begin{figure*}
\centering
   \includegraphics[width=18cm]{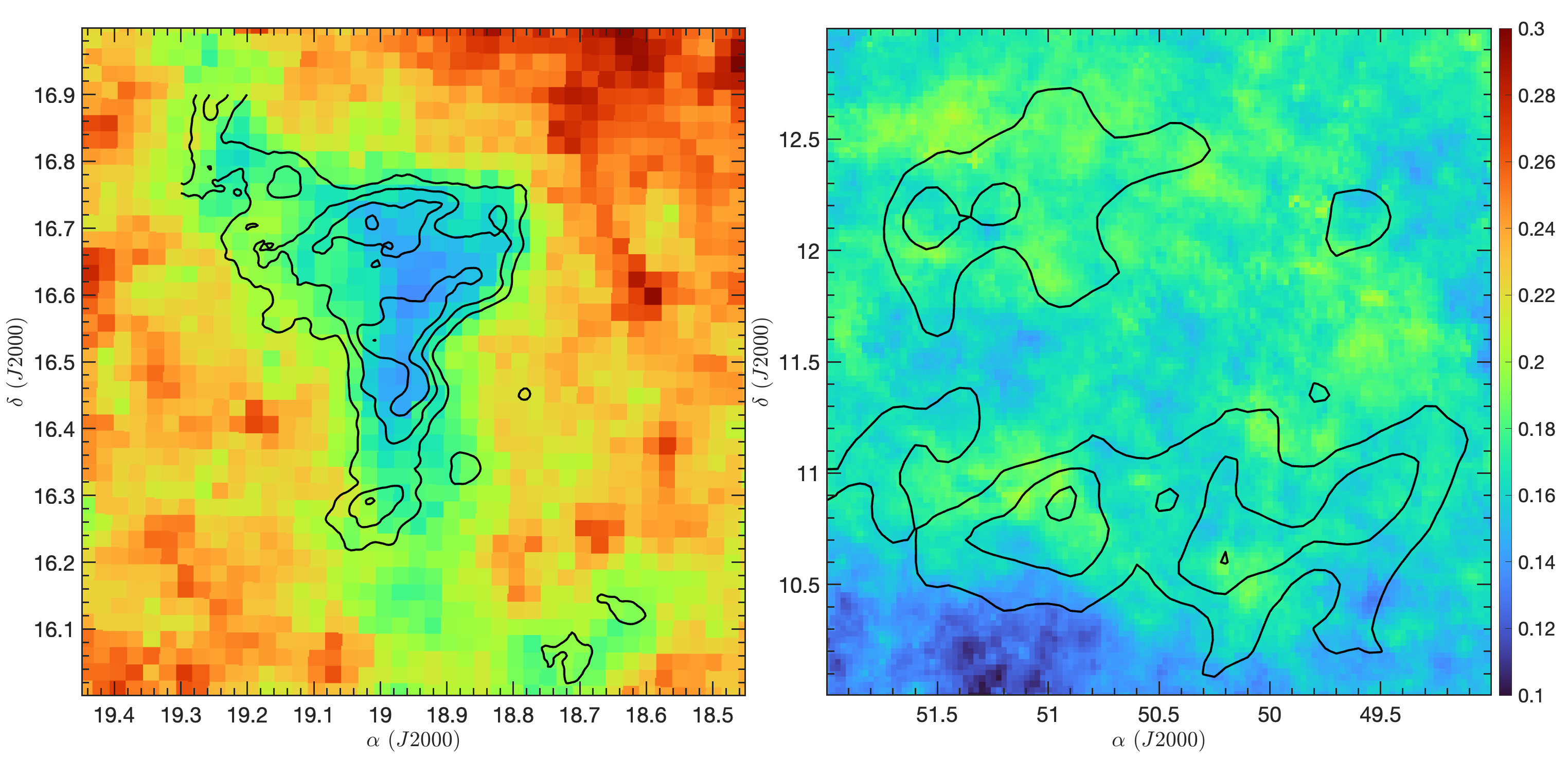}
     \caption{Ratio of 60 $\mu$m and 100 $\mu$m emission from IRIS survey for MBM 3 (left panel) and MBM 16 (right panel), that is a measure of dust temperature. The overplotted  $^{12}$CO (1-0) contours span between 3 and 12 K km s$^{-1}$ in steps of 3 K km s$^{-1}$ for MBM 3 and   1.5, 3, 4.5 K km s$^{-1}$ for MBM 16. The colorbar is the same for both maps.  }
     \label{fig:ratio_60_100}
\end{figure*}

\subsection{Dust temperatures}
The dust distribution and temperatures are different between MBM 3 and MBM 16, similar to the picture from  $^{12}$CO integrated antenna temperature data (see Figures \ref{fig:secondary_cloud} and \ref{fig:mbm16int}). The ratio between the 60 and 100 $\mu$m emission from IRIS survey, which is correlated with the dust temperature, is displayed in Figure \ref{fig:ratio_60_100}. The ratio clearly shows a region in MBM 3, which roughly traces the structure revealed by $^{12}$CO emission, systematically cooler than the environment. The mean Planck Dust GNILC model temperature \citep[see][]{2016A&A...596A.109P, 2020A&A...641A...4P} is $\sim 14$ K for the molecular core and $\sim 18$ K in the environment.  For  MBM 16, the entire region shows a mean temperature of $\sim 18$ K without distinguishing the molecular cloud. The MBM 3 result is similar to our conclusions for MBM 40 that these clouds tend to form in regions that are a few degrees lower than the environment. In contrast, for MBM 16, the mean inferred dust temperature within the $^{12}$CO contours is about the same as the environment, and the  60/100 $\mu$m ratio does not follow the $^{12}$CO contours.  This is not surprising given the complexity of the dust emission south of the Taurus dark clouds (see, e.g., \cite {1988LNP...306..168M}). The amount of foreground dust emission is likely to be significant in this direction, and identifying the dust specifically associated with MBM 16 will require a 3D dust map of the region similar to the work of \cite {2022A&A...661A.147L}.

\section{Conclusions}
From large-scale atomic structures extending over tens of parsecs to molecular concentrations within clouds, the ISM is organized in a broad spectrum of filaments.  Our analysis of three HLMCs (MBM 3 and MBM 16 in this paper, MBM 40 by \citep [] {2023A&A...676A.138M}) shows that these are present even without internal star formation and in the absence of self-gravity.  Velocity gradients found within these structures are not necessarily dynamical, convergent flows, and projection effects and topology of the driving flows produce signatures that mimic velocity shears even if they are simply distortions of ordered gas. Furthermore, depending on projection angles to the line of sight, the filaments can produce regions that have previously been interpreted as coherent structures.  Because of their relatively simple structures, the study of non-star-forming diffuse clouds may better elucidate some of the mechanisms responsible for the production of filaments in the diffuse ISM.

\nopagebreak

\begin{acknowledgements}
FCRAO was supported in part by the National Science Foundation under grant AST94-20159 and was operated with the permission of the Metropolitan District Commission, Commonwealth of Massachusetts.  This publication utilizes data from the Galactic ALFA HI (GALFA HI) survey data set obtained with the Arecibo L-band Feed Array (ALFA) on the Arecibo 305m telescope. At the time of the observations, the Arecibo Observatory was operated by SRI International under a cooperative agreement with the National Science Foundation (AST-1100968), and in alliance with Ana G. Méndez-Universidad Metropolitana, and the Universities Space Research Association. The GALFA HI surveys have been funded by the NSF through grants to Columbia University, the University of Wisconsin, and the University of California.

The Digitized Sky Survey was produced at the Space Telescope Science Institute under U.S. Government grant NAG W–2166. The images of these surveys are based on photographic data obtained using the Oschin Schmidt Telescope on Palomar Mountain and the UK Schmidt Telescope. The plates were processed into the present compressed digital form with the permission of these institutions.

This work has made use of data from the European Space Agency (ESA) mission
{\it Gaia} (\url{https://www.cosmos.esa.int/gaia}), processed by the {\it Gaia}
Data Processing and Analysis Consortium (DPAC,
\url{https://www.cosmos.esa.int/web/gaia/dpac/consortium}). Funding for the DPAC
has been provided by national institutions, in particular the institutions
participating in the {\it Gaia} Multilateral Agreement.

We warmly thank João Alves and Mario Tafalla for useful discussions, Henrik Olofsson and the Onsala Space Observatory for providing us the deep integration in CO to check the FCRAO data and for useful discussion about the reliability of the observation, Alex Kraus for obtaining new CH observations of MBM 3 with Effelsberg that anchored the velocity scale of the archival data,  Mark Heyer for his invaluable advice and assistance during our observations at FCRAO, and Nunzia Di Giacomo for her graphical magic.
\end{acknowledgements}

   \bibliographystyle{aa.bst} % style aa.bst
   \bibliography{biblio.bib} % your references Yourfile.bib

\begin{appendix}

\section{The mass of MBM 3}\label{app:mass}
The original total mass estimate, M$_{\mathrm{tot}}$ = M(\ion{H}{i} + H$_2$) = 44 M$_{\odot}$ \citep{2005PhDT.......331C}, only a quarter of our new estimate, was based on a distance of 130 pc. Since the \element[][12]CO dataset used in this paper is the same as \cite{2005PhDT.......331C}, the difference does not scale as the distance squared and must be due to the different assumptions and analysis of the \element[][12]CO data along with our different \ion{H}{i} dataset.  The \element[][12]CO antenna temperature from the FCRAO observations was recorded as T$_A^*$ after chopper wheel calibration \citep[see][]{1981ApJ...250..341K} that converts to the radiation temperature, T$_R$, by  T$_R = T_A^*/(\eta_{fss}\eta_c)$, where $\eta_{fss}$ is the scattering and spillover efficiency ($\approx$ 0.7 at 115 GHz at the time of the observations; Mark Heyer, private communication), and $\eta_c$ is the coupling factor of the antenna to the source, which \cite{2005PhDT.......331C} assumed to be unity. Alternatively, the main beam temperature (T$_{mb}$) is obtained by dividing the antenna temperature (T$_A^*$) by the beam efficiency ($\eta_{mb}$), which was $\approx$ 0.45 for the FCRAO telescope at 115 GHz at the time of the observations (Mark Heyer, private communication). In this paper, we converted the raw antenna temperature to T$_{mb}$ for all our mass determinations, while Chastain used the radiation temperature. Furthermore, as noted in \cite{1998ApJ...504..290M}, the X$_{CO}$ factor, which converts the total integrated antenna temperature to the H$_2$ column density, can change within a cloud and factors of 2-3 in the mass estimate may result solely from the choices of X$_{CO}$. 
 
 The mass estimation also depends on the velocity range chosen for the integration. For the molecular gas, the choice is generally well defined, since the lines only cover a few km s$^{-1}$. But for the atomic gas, the line profiles are much broader than the molecular lines, so the atomic mass estimate changes significantly depending on the velocity interval used. For example, if we integrate over the entire velocity range from $-$30 to 30 km s$^{-1}$, the atomic mass is approximately 60 M$_{\odot}$, but if we select channels corresponding to only the primary cloud, between $-$10.9 and 3.7 km$^{-1}$, the atomic gas mass estimate drops to 28 M$_{\odot}$. For our estimate, we used the \element[][12]CO as a benchmark, and we integrated both the atomic and molecular gas over the same velocity range.
 
 Finally, MBM 3 is surrounded by diffuse gas, which is traced by both \ion{H}{i} and \element[][12]CO observations. Since Chastain did not use any mask to select the region corresponding to MBM 3, his mass estimate may contain gas that is not directly linked with the cloud, yielding an overestimation of the total mass. Instead, we included only 21 cm emission above the 1 K km s$^{-1}$ level, thereby eliminating most of the surrounding diffuse atomic gas.
 
 Thus, the difference between our total mass estimate, 160 M$_\odot$, and the previous one, 44 M$_\odot$, is due to a combination of different velocity range selection, the use of main beam antenna temperature instead of radiation temperature, and the manner in which unrelated diffuse gas around the cloud was removed.

\section{Onsala 20-m telescope $^{12}$CO deep integration}\label{app:onsala_co}

 \cite{2010AJ....139..267C} discussed a possible velocity offset between CO and CH in MBM 40 and MBM 3. While a velocity offset seems likely for MBM 40, the offset in MBM 3 could not be confirmed since they claimed that a $+1.5$ km s$^{-1}$ velocity offset from the true LSR velocity was present in the  \element[][12]CO data presented by 
 \cite {2006A&A...457..197S}.  Since we use that dataset in this paper, we want to resolve this issue. Thus, we have independently checked the velocity scale of the FCRAO CO data for MBM 3 with a deep integration in \element[][12]CO using the Onsala 20-m telescope.

\begin{figure}
   \centering
   \includegraphics[width=\hsize]{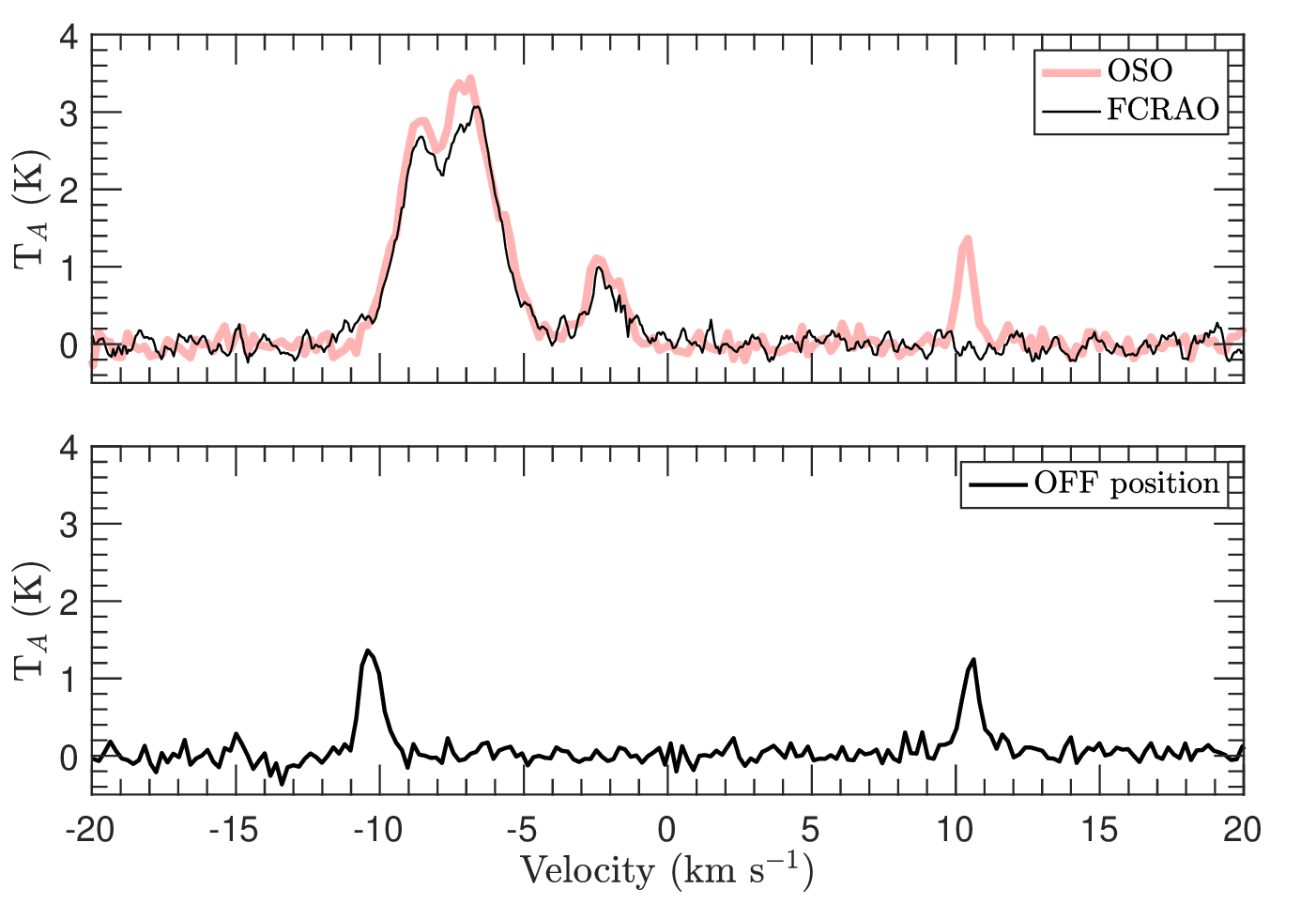}
     \caption{(Top panel). Onsala spectrum (thick light red line) and FCRAO (thin black line) of \element[][12]CO in Position 15. The line that shows at $+$10.5 km s$^{-1}$ is telluric in origin and should be ignored here. (Bottom panel). Onsala OFF (about 20\arcmin west from Position 15) spectrum used as a check for the telluric line. The line at $-$10.5 km s$^{-1}$ probably picks up some gas in the surroundings of the cloud. See the text for further details.}
         \label{fig:onsala_co_check}
 \end{figure}

The Onsala integration was made at Position 15 (see Figure \ref{fig:ch_pointings}) and
is superposed on the FCRAO \element[][12]CO spectrum for that location in Figure \ref{fig:onsala_co_check}.
The OFF spectrum is about 20\arcmin \ west of the on-source selected point. The telluric line is clearly visible in both Onsala spectra at about $+$10.5 km s$^{-1}$.
Aside from the slight difference in the peak antenna temperature between Onsala and FCRAO data, which could be an effect of focus or averaging through different observations done at different times at the Onsala Observatory, the match between the two spectra is excellent, which means that the FCRAO data, despite a lower SNR, have the correct LSR velocity scale.

The OFF position (lower panel) shows a line at about $-10.5$ km s$^{-1}$. Unfortunately, we cannot check its presence in the FCRAO data, since the OFF position is outside the mapped region. This spectral feature likely arises from low-level molecular gas at the outskirts of the cloud. Because atomic gas at the OFF position at that velocity is detected, it may be possible that a molecular component is also present.

\section{The artifact in \ion{H}{i} data near MBM 3 and MBM 16}\label{app:artifact}
\begin{figure}
  \centering
  \includegraphics[width=\hsize]{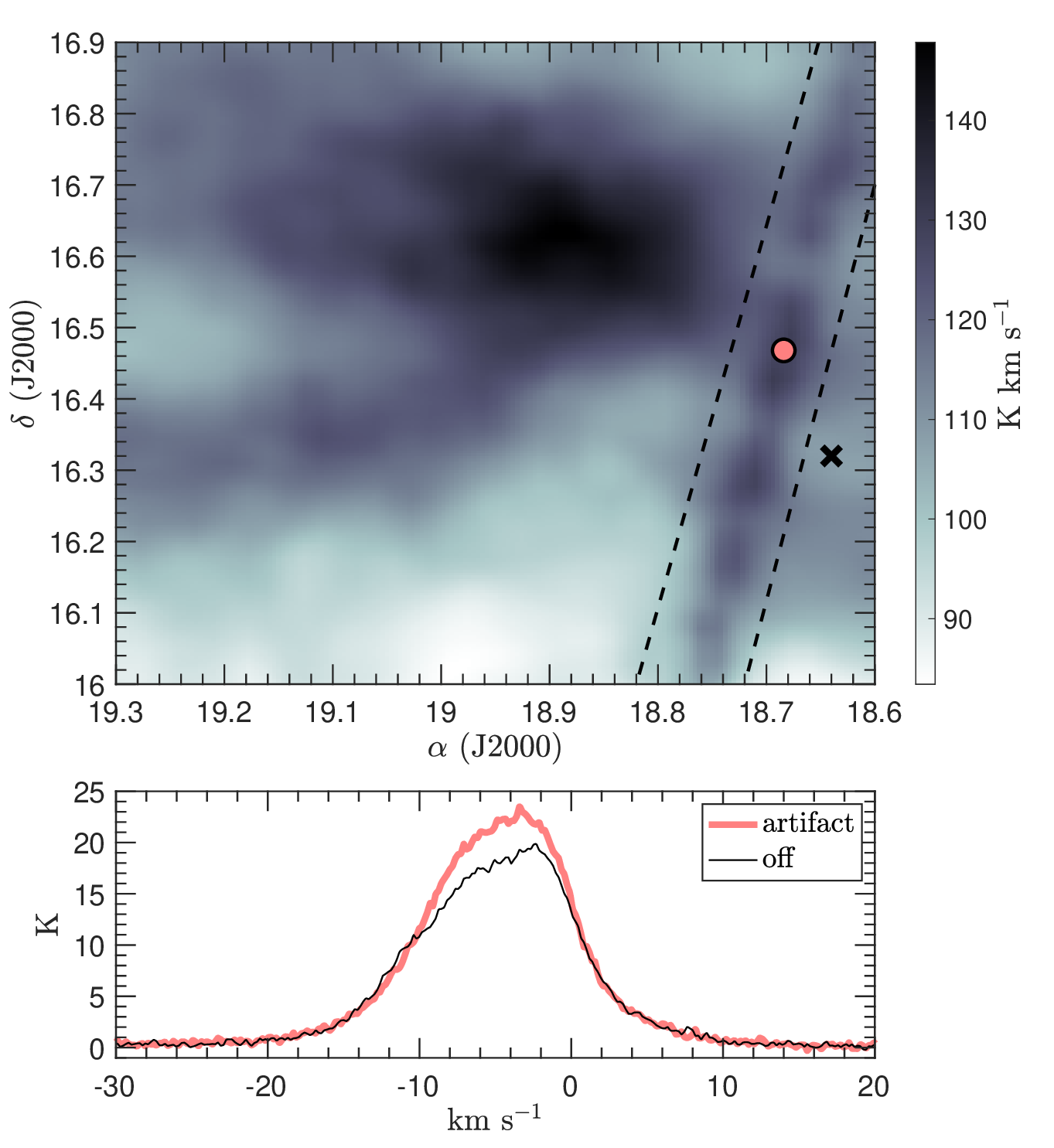}
    \caption{The ripple near MBM 3. The top panel shows the integrated \ion{H}{i} antenna temperature, and the two dashed lines bound the artifact, which appears as a series of knots with enhanced antenna temperature. The bottom panel displays two \ion{H}{i} spectra, one depicted by the thick orange line is within the artifact (the position is indicated in the top panel by an orange dot), and the other is outside the ripple, depicted by the thin black line in the bottom panel at the position highlighted by a black "x" in the top panel.}
        \label{fig:artifact}
\end{figure}

The GALFA data release 2 \citep{2018ApJS..234....2P} has, in general, much higher quality than the previous release, since the authors used more advanced reduction algorithms and paid careful attention to systematics. However, some residuals linger in the data, especially the ripples due to the pattern of the OTF mapping and the arrangement of detectors, which are clearly visible in their Figure 3 (bottom panel).

Unfortunately, two of these ripples appear near MBM 3 and MBM 16. Since the ripple near MBM 16 (see the velocity channels in Figure \ref{fig:vel_slices_mbm16}) is less pronounced in the velocity slices of interest, in this Appendix, we discuss only the one near MBM 3 (see Figure \ref{fig:artifact}).  Our conclusions for the impact of this ripple on our analysis conclusions will be the same for MBM 16.

To evaluate the role of the ripple in MBM 3, we carefully checked the spectra within this region to see if it introduced some systematics in our analysis. As reported in the bottom panel of Figure \ref{fig:artifact}, a sample spectrum shows enhanced antenna temperature in the velocity range of [$-$10, 0] km s$^{-1}$, but the line shape is similar to those outside the ripple. We can therefore conclude that our dynamical analysis is not affected since we use only the spectra within the cloud (in the right ascension range of [18.8\degr, 19.2\degr]), and so the atomic cloud mass is not altered by the presence of the ripple. Within the artifact, the atomic hydrogen column density N(\ion{H}{i}) is enhanced by 10\%.
As a last control, we used the Effelsberg-Bonn HI Survey \citep[EBHIS, see][]{2016A&A...585A..41W} data, checking sample positions within the cloud, within the ripple and outside the cloud where only the background gas is present. The EBHIS spectra show, in general, a peak antenna temperature about 10\% less than the GALFA in the same position, probably due to beam dilution since the Effelsberg telescope has a diameter of one-third with respect to the Arecibo telescope.

\end{appendix}

\end{document}